\documentclass[11pt, draftclsnofoot, journal, letterpaper, onecolumn]{IEEEtran}

\usepackage[dvips]{graphicx}
\usepackage{times}
\usepackage{cite}
\usepackage{amsmath}
\usepackage{array}
\usepackage{amssymb}

\usepackage{stfloats}
\usepackage{rotating,threeparttable,booktabs}
\usepackage{bm}
\usepackage{dcolumn,booktabs}
\usepackage{multirow}
\usepackage{graphicx}
\usepackage{subfigure}
\usepackage{xcolor}
\begin{document}

\title{\huge {Optimal Linear Transceiver Designs for Cognitive Two-Way Relay Networks}}

\author{\IEEEauthorblockN{Rui~Wang, Meixia Tao, \IEEEmembership{Senior Member,~IEEE}, and Yuan Liu, \IEEEmembership{Student Member, IEEE}}
\thanks{Copyright (c) 2011 IEEE. Personal use of this material is permitted. However, permission to use this material for any other purposes must be obtained from the IEEE by sending a request to pubs-permissions@ieee.org.}
\thanks{The authors are with the Department of Electronic Engineering at Shanghai Jiao Tong University, Shanghai, 200240,
P. R. China. Emails:\{liouxingrui, mxtao, yuanliu\}@sjtu.edu.cn.}
\thanks{This work is supported by the
National 973 project under grant 2012CB316100,
the National Natural Science Foundation of China under grant 60902019 and
New Century Excellent Talents in University (NCET) under grant NCET-11-0331.}}

\maketitle
\vspace{-1cm}
\begin{abstract}
This paper studies a cooperative cognitive radio network where two primary users (PUs) exchange information with the help of a secondary user (SU) that is equipped with multiple antennas and in return, the SU superimposes its own messages along with the primary transmission.
The fundamental problem in the considered network is the design of transmission strategies at the secondary node. It involves three
basic elements: first, how to split the power for relaying the primary signals and for transmitting the secondary signals; second, what two-way relay strategy should be used to assist the bidirectional communication between the two PUs;  third, how to jointly design the primary and secondary transmit precoders.
This work aims to
address this problem by proposing a transmission framework of
maximizing the achievable rate of the SU while maintaining the rate requirements of the two PUs. Three well-known and practical two-way relay strategies are considered:
amplify-and-forward (AF), bit level XOR based decode-and-forward (DF-XOR) and symbol level
superposition coding based DF (DF-SUP).
For each relay strategy, although the design problem is non-convex,
we find the optimal solution by
using certain transformation techniques and optimization tools such as semidefinite programming (SDP) and second-order cone programming (SOCP).
Closed-form solutions are also obtained under certain conditions.
Simulation results show that when the rate requirements of the two PUs are symmetric,
by using the DF-XOR strategy
and applying the proposed optimal precoding, the SU requires the least power for relaying and thus reserves the most power to transmit its own signal.
In the asymmetric scenario, on the other hand,
the DF-SUP strategy with the corresponding optimal precoding
is the best.

\end{abstract}


\section{Introduction}
Due to the increasing popularity of wireless devices,
the radio spectrum has been
an extremely scarce resource. By contrast,
most of the existing licensed spectrum remains under-utilized.
Cognitive radio (CR) is an efficient way
to improve spectrum utilization \cite{Haykin05, Goldsmith2009}.
The basic idea of CR is to allow unlicensed or secondary users (SUs) to access the licensed spectrum originally allocated to primary users (PUs)
without sacrificing the quality-of-service (QoS) of the PUs.
Some fundamental problems, such as reliable spectrum sensing \cite{Wang2010} and dynamical spectrum access (see \cite{QingZhao2007} and the reference therein), have been well studied.
Recently, combining CR with cooperative or relay techniques has received a great deal of interest from both academia and industry since it can make CR more reliable in application \cite{Musavian2010,Han2009,YonghuiLi2011,YangHan2010}.
It is worth noting that most of these existing works focus on unidirectional communications using traditional one-way relay strategies.

Due to bidirectional or two-way nature of communication networks,
a promising relay technique, two-way relaying, has been proposed recently.
Two-way relaying applies the principle of physical layer network coding (PLNC) at the relay node so as to mix the signals received from the two source nodes, and then employs self-interference (SI) cancelation at each destination to extract the desired information \cite{Rankov2007, Zhang2009,RuiWang,WangRui2011,Yuan2010}. As a result, two-way relaying needs less time slots to complete information exchange between two sources
and has higher spectral efficiency than the traditional one-way relaying.
It is thus natural to incorporate two-way relaying into CR networks to further enhance the spectrum utilization. One possible scenario is to apply dedicated relay nodes to assist the bidirectional communication of secondary networks as in \cite{Jitvanichphaibool2009,Safavi2011}.
In specific, authors in \cite{Jitvanichphaibool2009} considered the two-way relaying between a pair of SUs with a dedicated multi-antenna amplify-and-forward (AF) relay node, and studied the problem of joint beamforming and power allocation with interference constraint at the PU. Authors in \cite{Safavi2011}
considered a similar network model but with
multiple dedicated single-antenna AF relays, and investigated the distributed beamforming design at the secondary network to minimize interference at the PUs with the SUs' signal-to-interference-plus-noise ratio (SINR) constraints.

In this work, we consider a different transmission protocol where
users in the primary network conduct bidirectional communication with the help of
a multi-antenna secondary node, rather than dedicated relay nodes.
Specifically, the multi-antenna secondary node acts as a relay to help the information exchange between two PUs,
and as a return, the secondary node is allowed to
simultaneously send its own messages in the same frequency band to the secondary receiver.
The considered protocol can be viewed as an overlay model \cite{Goldsmith2009}, which creates a ``win-win" situation for both PUs and SUs. Under this setting, two primary signals should be first combined together via physical layer network coding at the secondary node, and then superimposed with the secondary signal.
Three issues should be carefully treated in the design of transmission strategies at the secondary node, including 1) how to split the power for relaying the primary signals and for transmitting the secondary signals; 2) what two-way relay strategy should be used to assist the bidirectional communication between the two PUs; and 3) how to jointly design the primary and secondary transmit precoders.

Note that using two-way relaying to assist primary transmission has also been considered in works \cite{Alizadeh2010,Li2011a}. Specifically, authors in \cite{Alizadeh2010} studied the beamforming design at the secondary transmitters for minimizing the total system power while guaranteeing the SINR requirements of all receivers. However, in \cite{Alizadeh2010}, the secondary transmitters exclusively act as AF relays when the PU pair is active or transmit their own signals only when the PU pair is inactive. Authors in \cite{Li2011a} considered a similar overlay  protocol as ours. However, it focused on outage performance analysis for a three-phase single-antenna CR network with bit level XOR based decode-and-forward (DF-XOR) relay strategy.

In this paper, we consider a two-phase overlay cognitive two-way relay network.
In the first phase, two PUs transmit their signals to a multi-antenna secondary node simultaneously. In the second phase,
after combining the two primary signals using physical layer network coding,
the secondary node superimposes its own message
and then broadcasts the resulting signal to the two primary receivers as well as its own secondary receiver.
We aim to address the aforementioned three issues, namely,
relay strategy selection, power splitting and joint precoding design by
proposing a transmission framework of
maximizing the achievable rate of the SU while maintaining the rate requirements of the two PUs. To achieve this goal,
we first identify three popular and practical two-way relay strategies: AF, DF-XOR and symbol level superposition coding based DF (DF-SUP). Then, for each relay strategy we find the optimal power splitting and joint precoding design at the secondary node.
It is shown that each design problem is non-convex.
By transforming these problems into more tractable forms, some efficient optimization tools, such as semidefinite programming (SDP) and second-order cone programming (SOCP), are applied to find the optimal solutions of all the schemes. Moreover, we derive the optimal closed-form solutions in several cases
where some of the channels are parallel in the second phase.
Simulation results show that when the rate requirements of the two PUs are symmetric,
by using the DF-XOR strategy and applying the proposed optimal precoding, the SU requires the least power for relaying and thus reserves the most power to transmit its own signal.
However, when the rate requirements of the two PUs are asymmetric, the DF-SUP relay strategy with the corresponding optimal precoding is the best and requires the least relay power consumption in satisfying the rate requirements of the PUs.

The rest of this paper is organized as follows. In Section II, the cognitive two-way relay system model is described.
Solving associated optimization problems by using suitable optimization tools is presented in Section III.
Extensive simulation results are illustrated in Section IV. Finally, Section
VI offers concluding remarks.

\emph{Notations}: $\cal E(\cdot)$ denotes the expectation over the
random variables within the bracket.
$\otimes$ denotes the Kronecker operator.
Superscripts $(\cdot)^T$, $(\cdot)^{*}$ and $(\cdot)^H$ denote the transpose, conjugate and conjugate transpose, respectively.
${\rm Tr}({\bf A})$, ${\bf A}^{-1}$ $\det({\bf A})$ and ${\rm Rank}(\bf A)$ stand for the trace, inverse, determinant and the rank of matrix ${\bf A}$, respectively.
${\rm Diag}(\bf a)$ denotes a diagonal matrix with ${\bf a}$ being its diagonal entries.
${\bf 0}_{N\times M}$ implies the
$N\times M$ zero matrix and ${\bf I}_N$ denotes the $N \times N$ identity matrix.
$|z|$ implies the norm of the complex number $z$, ${\Re}(z)$ and $\Im (z)$ denote the real and imaginary part of $z$, respectively.
$||{\bf x}||^2_2$ denotes the squared Euclidean norm of a complex vector ${\bf x}$ and $||{\bf X}||^2_F$ denotes the Frobenius norm of a complex matrix ${\bf X}$. The distribution of a circular symmetric complex Gaussian vector with mean vector $\bf x$ and covariance matrix ${\bf \Sigma} $ is denoted by ${\cal CN}({\bf x},{\bf \Sigma})$.
${\mathbb C}^{x \times y}$ denotes the space of $x \times y$ matrices with complex entries.

\section{System Model}
\begin{figure}[tbhp]
\begin{centering}
\includegraphics[scale=0.65]{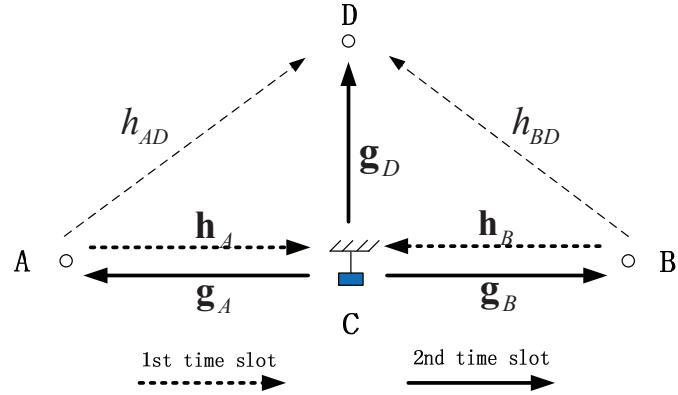}
\vspace{-0.1cm}
\caption{Illustration of a cognitive two-way relay network.} \label{Fig0}
\end{centering}
\vspace{-0.3cm}
\end{figure}

Consider a primary network, where
two PUs, denoted as $A$ and $B$, intend to exchange information in a licensed frequency band as shown in Fig.~\ref{Fig0}.
Due to impairments such as multipath fading, shadowing, path loss of wireless channels and obstacles etc., the direct communication channel between $A$ and $B$ is assumed not strong enough to support a target data rate for information exchange.
They thus seek cooperation with a nearby node $C$ from the secondary network.
That is to say, the secondary node $C$ acts as a relay to assist the bidirectional communication between $A$ and $B$.
As a return, the secondary node $C$ is allowed to superimpose its own message
into the relayed primary signals and then
broadcasts the resulting signal to the two primary receivers as well as its own secondary receiver $D$.

Due to the absence of direct link, we assume that two-phase two-way relaying protocol is employed to complete the bidirectional communication.
Specifically, in the first phase (also referred as multiple access (MAC) phase), both $A$ and $B$ transmit their signals to the secondary node $C$ simultaneously.
By assuming that $M$ antennas are equipped at $C$, the received signal vector at $C$ is denoted as
\begin{equation} \label{I-eq1} \nonumber
{\bf y}_C = {\bf h}_A  s_A + {\bf h}_B  s_B + {\bf n}_C,
\end{equation}
where $s_i$, for $i\in \{A,B\}$, represents the transmit signal from the PU $i$.
${\bf h}_i \in {\mathbb C}^{M \times 1}$ is the channel vector from the PU $i$ to the secondary node $C$, and ${\bf n}_C$ denotes the additive complex Gaussian noise vector at $C$ following ${\cal CN}({\bf 0},\sigma^2_C {\bf I}_{M})$.
Each transmit signal $s_i$ is assumed to satisfy an average power constraint, i.e., ${\cal E}(|{s}_i|^2)=P_i$.
In the meantime, the secondary receiver $D$, equipped with single antenna, can also overhear
the signals from the PUs $A$ and $B$, and the received signal is given by
\begin{equation} \label{I-eq2} \nonumber
y_{D,1} =  h_{AD}  s_A +  h_{BD}  s_B + n_{D,1},
\end{equation}
where $h_{AD}$ and $h_{BD}$ denote the channel gains from the PUs $A$ and $B$, respectively, to the secondary receiver $D$, and $n_{D,1}$ denotes the additive complex Gaussian noise at $D$ following ${\cal CN}(0,\sigma^2_D )$. The secondary receiver $D$ can decode the received signals in the first phase,
which can be treated as side information for improving the performance of the secondary transmission in the second phase.

Upon receiving ${\bf y}_C$, the secondary node $C$ performs certain processing
and then forwards it together with its own message in the second phase, also referred as broadcast (BC) phase.
Let the transmit signal from $C$ be denoted as
\begin{equation} \label{I-eq2-1}
{\bf x}_C =  {\bf x}_{AB} + {\bf s},
\end{equation}
where ${\bf x}_{AB}$ is the combined signal of the two primary messages by using PLNC and ${\bf s}$ denotes the signal intended
to the secondary receiver $D$. As mentioned earlier, the fundamental problem here is to design the structures of ${\bf x}_{AB}$ and ${\bf s}$, and the power splitting between them.

By adopting different two-way relay strategies, the transmit signal ${\bf x}_{AB}$ can be different. In the case of pure two-way relaying (i.e., ${\bf s}={\bf 0}$), the optimal design of ${\bf x}_{AB}$ is essentially equivalent to designing a coding strategy to achieve multiple single-link capacities in BC phase by transmitting one encoded signal as in \cite{Oechtering2008IT, Kramer2007, Sang2008}. Intuitively, the optimal relay strategy in our considered network should be also designed like this. However, in this work we only focus on using some sub-optimal relay strategies since the capacity-achieving two-way relay strategies proposed in \cite{Oechtering2008IT, Kramer2007, Sang2008} are derived from the information theoretic perspective and hence require techniques such as random binning and jointly typical set decoding which are difficult to realize in practice. The primary focus of this work is to obtain the optimal and specific linear precoding structure based on practical two-way relay strategies. The three sub-optimal strategies we considered, namely, AF, DF-XOR and DF-SUP, are all favorable for practical implementation and the precoding designs based on these strategies are mathematically tractable.

\subsection{AF Relay Strategy}
By applying AF relay strategy, the signal ${\bf x}_{AB}$ for the PUs in \eqref{I-eq2-1} can be expressed as
\begin{equation} \label{I-eq3} \nonumber
{\bf x}_{AB} = {\bf W} {\bf y}_C = {\bf W} {\bf h}_A s_A + {\bf W} {\bf h}_B s_B +  {\bf W} {\bf n}_C,
\end{equation}
where ${\bf W}$ represents the precoding matrix for the primary signals. In addition, we assume that the secondary node $C$ has the maximum transmit power $P_C$, which yields
\begin{equation} \label{I-eq4}
{\cal E}({\bf x}_C {\bf x}^H_C ) = P_A ||{\bf W} {\bf h}_A ||^2_2 + P_B||{\bf W} {\bf h}_B ||^2_2 + {\rm Tr}({\bf Q}_s) + {\sigma^2_C} ||{\bf W}||^2_F \leq P_C,
\end{equation}
where ${\bf Q}_s = {\cal E} ({\bf s} {\bf s}^H)$ is the covariance matrix of ${\bf s}$. Then the received signals at $A$ and $B$ are given by
\begin{equation} \label{I-eq5}
\tilde{y}_i = {\bf g}^T_i {\bf W} {\bf h}_{\bar{i}} s_{\bar{i}} +  {\bf g}^T_i {\bf W} {\bf h}_i s_i  + {\bf g}^T_i {\bf s} + {\bf g}^T_i {\bf W} {\bf n}_C +n_i,~i=A,B
\end{equation}
where ${\bar i}=B$ if $i=A$ and ${\bar i}=A$ if $i=B$, ${\bf g}_i$ denotes the channel vector from the secondary node $C$ to the destination node $i$, and $n_i$ denotes the additive Gaussian noise at the destination node $i$ following ${\cal CN}(0,\sigma^2_i)$ for $i\in \{A,B\}$.
The received signal at the secondary receiver $D$ in the second phase is given by
\begin{equation} \label{I-eq7}
y_{D,2} = {\bf g}^T_D {\bf W} {\bf h}_A s_A +  {\bf g}^T_D {\bf W} {\bf h}_B s_B  + {\bf g}^T_D {\bf s} + {\bf g}^T_D {\bf W} {\bf n}_C +n_{D,2},
\end{equation}
where ${\bf g}_D$ represents the channel vector from the secondary node $C$ to the secondary receiver $D$, and $n_{D,2}$ denotes the additive Gaussian noise at $D$ in the second phase following ${\cal CN}(0,\sigma^2_D)$.
Since the PUs $A$ and $B$ know their own transmit messages $s_A$ and $s_B$ a prior, respectively, the back propagated self-interference term $s_i$ can be subtracted from \eqref{I-eq5} before demodulation. The equivalent received signals at $A$ and $B$ are thereby yielded as
\begin{equation} \label{I-eq8}
y_i =  {\bf g}^T_i {\bf W} {\bf h}_{{\bar i}} s_{{\bar i}}  + {\bf g}^T_i {\bf s} + {\bf g}^T_i {\bf W} {\bf n}_C+n_i,~i=A,B.
\end{equation}
Similarly, if the secondary receiver $D$ can decode $s_A$ or/and $s_B$, the corresponding interference can be subtracted from \eqref{I-eq7}, which is helpful for improving the performance of the secondary transmission. The details shall be discussed in the next section.

\subsection{DF-XOR Relay Strategy}
If the secondary node $C$ adopts DF relay strategy, namely DF-XOR and DF-SUP\cite{Rankov2007,Oechtering2008}, it needs to decode the received signals in the first phase, which is known as a MAC channel. We assume that the secondary node $C$ has enough processing ability to correctly decode the received signals if the transmit rates from the two PUs lie in the rate region given as follows
\begin{equation} \label{I-eq10}
\mathcal{C}_{MAC} (R_A, R_B)= \left\{ \begin{array}{cc}
                              &R_A \leq \log_2 \left(1+\frac{P_A ||{\bf h}_A||^2_2}{\sigma^2_C} \right)~~~~~~~~~~~~~~~  \\
                              &R_B \leq \log_2 \left(1+\frac{P_B ||{\bf h}_B||^2_2}{\sigma^2_C} \right)~~~~~~~~~~~~~~~ \\
                              &R_A+R_B \leq \log_2 \det\left({\bf I}_M +\frac{P_A}{\sigma^2_C} {\bf h}_A {\bf h}^H_A+ \frac{P_A}{\sigma^2_C} {\bf h}_B {\bf h}^H_B \right)
                            \end{array}
                            \right.
\end{equation}
where $R_A$ and $R_B$ are the transmit rates of the PUs $A$ and $B$, respectively. If any of $s_A$ and $s_B$ has not been correctly decoded, we claim that the primary transmission is in outage.

Let ${\bf b}_i$ denote the decoded bit sequence from $s_i$, for $i \in \{A,B\}$. By applying XOR operation, the combined bit sequence is yielded as ${\bf b}_{AB}={\bf b}_A \oplus {\bf b}_B$\footnote{If the lengths of the bit sequences ${\bf b}_A$ and ${\bf b}_B$ are different, zero-padding is exploited to the shorter one to make it have the same length as the longer one.} where $\oplus$ denotes the XOR operator. Then the combined bit sequence ${\bf b}_{AB}$ is encoded and modulated as an $M \times 1$ signal ${\bf s}_{AB}$.
Thus we have ${\bf x}_{AB}={\bf s}_{AB}$.
To satisfy the power constraint at $C$, we have
\begin{equation} \label{I-eq10-1}
{\cal E}({\bf x}_C {\bf x}^H_C ) =  {\rm Tr}({\bf Q}_{AB}) + {\rm Tr}({\bf Q}_s) \leq P_C,
\end{equation}
where ${\bf Q}_{AB} = {\cal E} ({\bf s}_{AB} {\bf s}^H_{AB})$ is the covariance matrix of ${\bf s}_{AB}$.
The received signal at each primary destination is given by
\begin{equation} \label{I-eq11}
y_i =  {\bf g}^T_i {\bf s}_{AB}  + {\bf g}^T_i {\bf s} + n_i,~i=A,B.
\end{equation}
Each PU $i$ can demodulate the received signal $y_i$ and then XOR it with its own transmit bits to obtain the desired information.
Similarly, the received signal at the secondary receiver $D$ is given by
\begin{equation} \label{I-eq13}
y_{D,2} = {\bf g}^T_D  {\bf s}_{AB} + {\bf g}^T_D {\bf s} + n_{D,2}.
\end{equation}
If $D$ correctly decodes both $s_A$ and $s_B$ in the first phase, the interference term ${\bf g}^T_D  {\bf s}_{AB}$ can be subtracted from $\eqref{I-eq13}$.

\subsection{DF-SUP Relay Strategy}
If the secondary node $C$ adopts the DF-SUP relay strategy,
we have ${\bf x}_{AB}={\bf s}_A + {\bf s}_B$,
where ${\bf s}_i$, for $i\in \{A,B\}$, represents the re-encoded and modulated signal of the PU $i$.
The power constraint at $C$ is then denoted as
\begin{equation} \label{I-eq13-1}
{\cal E}({\bf x}_C {\bf x}^H_C ) =  {\rm Tr}({\bf Q}_A) +{\rm Tr}({\bf Q}_B)+ {\rm Tr}({\bf Q}_s) \leq P_C,
\end{equation}
where ${\bf Q}_i = {\cal E} ({\bf s}_i {\bf s}^H_i)$, for $i\in \{A,B\}$, is the covariance matrix of ${\bf s}_i$.
After self-interference cancelation, the received signal at each primary destination is yielded as
\begin{equation} \label{I-eq14}
y_i =  {\bf g}^T_i {\bf s}_{\bar{i}} + {\bf g}^T_i {\bf s} + n_i,~i=A,B.
\end{equation}
The received signal at the secondary receiver $D$ is denoted as
\begin{equation} \label{I-eq16}
y_{D,2} = {\bf g}^T_D ({\bf s}_A + {\bf s}_B) + {\bf g}^T_D {\bf s} + n_{D,2}.
\end{equation}
Here, any correctly decoded message in the first phase can be applied to subtract the corresponding interference in \eqref{I-eq16} as in the AF case.

Before leaving this section, we provide some discussions on the cooperation between the PUs and the SU in the considered cognitive two-way relay network.
In this work, we assume that all the designs are performed at the secondary node $C$ and thus following network channel state information (CSI) are needed at $C$.
The channel vectors ${\bf h}_A$ and ${\bf h}_B$ can be measured by $C$ itself.
The channel vectors ${\bf g}_A$ and ${\bf g}_B$ in the reverse links
can be measured and sent by $A$ and $B$, respectively, via
a feedback channel to $C$\footnote{Here we assume that the PUs are cooperative and feed back ${\bf g}_A$ and ${\bf g}_B$ correctly. This assumption is widely used in the literatures \cite{Safavi2011, Jitvanichphaibool2009, Li2011a}.}.
If channel reciprocity holds (for example in time-division duplex systems),
we have ${\bf g}_A={\bf h}_A$ and ${\bf g}_B={\bf h}_B$,
and thus no CSI feedback is needed for nodes $A$ and $B$.
Note that the channels $h_{AD}$ and $h_{BD}$ are not needed at $C$,
and $C$ only needs the secondary receiver $D$ to report
whether it correctly decodes the PUs' signals in the first phase or not. This message is also local with respect to the secondary receiver $D$ and the secondary node $C$.
Thus we claim that in our considered cognitive two-way relay network, the optimization at $C$ only needs local information and is applicable in practical systems.

\section{Linear Transceiver Designs}
In this section, linear transceiver designs at the secondary node $C$ associated with different relay strategies are considered. Our objective is to maximize the achievable rate of the SU while maintaining the rate requirements of the two PUs. Note that the power splitting is embedded in the transceiver design automatically and will not be discussed separately in this section.

\subsection{Joint Design of ${\bf W}$ and ${\bf Q}_s$ Under AF Two-Way Relay Strategy}
Based on \eqref{I-eq7}, \eqref{I-eq8}, the achievable rates of the PUs and the SU are denoted, respectively, as
\begin{equation} \label{II-1-eq1}
\begin{split}
\gamma^{AF}_i &= \frac{1}{2}\log_2 (1+ {\rm SINR}^{AF}_i), ~i=A,B,\\
\gamma^{AF}_D &= \frac{1}{2}\log_2 (1+ {\rm SINR}^{AF}_D).
\end{split}
\end{equation}
Here the factor $1/2$ results from the fact that two phases are required for the cooperative transmission. The SINRs in \eqref{II-1-eq1} are given, respectively, by
\begin{equation} \label{II-1-eq3} \nonumber
{\rm SINR}^{AF}_i = \frac{P_{\bar i} |{\bf g}^T_i {\bf W} {\bf h}_{\bar i} |^2}{{\bf g}^T_i {\bf Q}_s {\bf g}^*_i + \sigma^2_C ||{\bf g}^T_i {\bf W} ||^2_2 +\sigma^2_i }, ~i=A,B
\end{equation}
and
\begin{equation} \label{II-1-eq4} \nonumber
{\rm SINR}^{AF}_D = \frac{{\bf g}^T_D  {\bf Q}_s {\bf g}^*_D }{ a_A P_A |{\bf g}^T_D {\bf W} {\bf h}_A|^2 + a_B P_B|{\bf g}^T_D {\bf W} {\bf h}_B|^2 + \sigma^2_C ||{\bf g}^T_D {\bf W} ||^2_2 + \sigma^2_D},
\end{equation}
where $a_i$, for $i\in \{A,B\}$, is a binary indictor with $a_i=0$ indicating that the secondary receiver $D$ correctly decodes the signal from the PU $i$ and the corresponding interference is then subtracted from the received signal in \eqref{I-eq7} and otherwise $a_i=1$. The optimization problem is thus yielded as
\begin{eqnarray} \label{II-1-eq5}
      \max_{{\bf W},{\bf Q}_s\succeq 0 } && {\rm SINR}^{AF}_D \\ \nonumber
      s.t. && {\rm SINR}^{AF}_i \geq \tau_i,~i=A,B \\ \nonumber
     && P_A ||{\bf W} {\bf h}_A ||^2_2 + P_B||{\bf W} {\bf h}_B ||^2_2 + {\rm Tr}({\bf Q}_s) + {\sigma^2_C} ||{\bf W}||^2_F \leq P_C
\end{eqnarray}
where $\tau_i = 2^{2 R_{\bar{i}}} - 1$, for $i \in  \{A,B\}$, with $R_i$ denoting the rate requirement of the PU $i$. To proceed to solve \eqref{II-1-eq5}, we have the following lemma.

\textbf{Lemma} 1: The optimal ${\bf Q}_s$ in \eqref{II-1-eq5} can be rank-one and denoted as ${\bf Q}_s = \bar{{\bf q}} \bar{{\bf q}}^H$, where the optimal $\bar{{\bf q}}$ has the form as $\bar{{\bf q}} = {\bf U} {\bf q}$. Here ${\bf q} \in \mathbb{C}^{N \times 1}$, ${\bf U}=[{\bf u}_1, \cdots, {\bf u}_N ]\in \mathbb{C}^{M \times N}$ ($N \leq 3$) with $\{ {\bf u}_1, \cdots, {\bf u}_N \}$ being the orthonormal bases which span space $\mathbb{G} = \{{\bf g}^*_D, {\bf g}^*_A, {\bf g}^*_B \}$.
\begin{proof}
Please refer to Appendix~\ref{prof_lemma1}.
\end{proof}

\emph{Lemma 1} indicates that for the secondary signal ${\bf s}$ at $C$, the beamforming is indeed optimal. Moreover, the not-yet-determined elements in ${\bf q}$ are irrelevant to the relay antenna number $M$ and only depend on $N$, i.e., the dimension of space $\mathbb{G}$. Thus, the computational complexity can be reduced in solving \eqref{II-1-eq5}. Based on \emph{Lemma 1}, optimization problem \eqref{II-1-eq5} is simplified as follows
\begin{subequations}\label{II-1-eq6}
\begin{align}
    \max_{{\bf W}, {\bf q}}  ~& {\rm SINR}^{AF}_D = \frac{|{\bf t}_D {\bf q}|^2}{ a_A P_A|{\bf g}^T_D {\bf W} {\bf h}_A|^2 + a_B P_B|{\bf g}^T_D {\bf W} {\bf h}_B|^2 + \sigma^2_C ||{\bf g}^T_D {\bf W} ||^2_2 + \sigma^2_D} \label{II-1-eq6-1}\\
    s.t.~  &  \frac{P_{\bar i} |{\bf g}^T_i {\bf W} {\bf h}_{\bar i} |^2}{|{\bf t}_i {\bf q}|^2 + \sigma^2_C ||{\bf g}^T_i {\bf W} ||^2_2 +\sigma^2_i } \geq \tau_i, ~i=A,B \label{II-1-eq6-2}\\
    & {\rm Tr}\left\{ {\bf W}(P_A {\bf h}_A{\bf h}^H_A + P_B {\bf h}_B{\bf h}^H_B + \sigma^2_C {\bf I}_M ){\bf W}^H \right\} + {\rm Tr}\left\{ {\bf q}{\bf q}^H\right\} \leq P_C \label{II-1-eq6-3}
\end{align}
\end{subequations}
where ${\bf t}_D = {\bf g}^T_D {\bf U}$, ${\bf t}_A = {\bf g}^T_A {\bf U}$ and ${\bf t}_B = {\bf g}^T_B {\bf U}$, inequality \eqref{II-1-eq6-3} is obtained by reformulating the power constraint in \eqref{I-eq4}. It is not hard to verify that optimization problem \eqref{II-1-eq6} is non-convex.
Next, we will find the optimal solution of this non-convex problem.

We first rewrite the objective function in \eqref{II-1-eq6-1} into the form as
\begin{equation}\label{Add-1}
\begin{split}
 {\rm SINR}^{AF}_D =& \frac{|{\bf t}_D {\bf q}|^2}{ a_A P_A|{\bf g}^T_D {\bf W} {\bf h}_A|^2 + a_B P_B|{\bf g}^T_D {\bf W} {\bf h}_B|^2 + \sigma^2_C ||{\bf g}^T_D {\bf W} ||^2_2 + \sigma^2_D}  \\
 =& \frac{{\rm Tr}( {\bf Q}_{01} {\bf q} {\bf q}^H) } {  {\bf w}^H {\bf Q}_{02} {\bf w} +\sigma^2_D}
\end{split}
\end{equation}
where ${\bf w}=vec({\bf W})$, ${\bf Q}_{01} = {\bf t}^H_D {\bf t}_D$ and
\begin{equation}\label{Add-2}
{\bf Q}_{02} = \left( a_A P_A {\bf h}_A {\bf h}^H_A + a_B P_B {\bf h}_B {\bf h}^H_B + \sigma^2_C {\bf I}_M  \right)^T \otimes \left({\bf g}^{*}_D {\bf g}^T_D \right).
\end{equation}
Equation \eqref{Add-1} is acquired by using the rule \cite{xiandazhang2004}
\begin{equation}\label{Add-3}
     {\rm Tr}\left( {\bf A} {\bf B} {\bf C} {\bf D}\right)=\left(vec({\bf D}^T)\right)^T \left({\bf C}^T \otimes {\bf A}\right) vec({\bf B}),
\end{equation}
then we get ${\bf Q}_{02}$ given in \eqref{Add-2}. Similar to \eqref{Add-1}, we can also transform the SINR constraint \eqref{II-1-eq6-2} into the form as
\begin{equation}\label{Add-4}
\begin{split}
      \frac{ {\bf w}^H {\bf Q}^i_1 {\bf w}  } {{\rm Tr}({\bf Q}^i_2 {\bf q} {\bf q}^H)+  {\bf w}^H {\bf Q}^i_3 {\bf w} + \sigma^2_i} \geq \tau_i,
\end{split}
\end{equation}
where ${\bf Q}^i_1 = P_{\bar i}({\bf h}_{\bar i} {\bf h}^H_{\bar i} )^T \otimes ({\bf g}^*_i {\bf g}^T_i  )$, ${\bf Q}^i_2 = {\bf t}^H_i {\bf t}_i$ and ${\bf Q}^i_3 = \sigma^2_C {\bf I}_M \otimes ({\bf g}^*_i {\bf g}^T_i  )$. Again by using \eqref{Add-3}, the power constraint in \eqref{II-1-eq6-3} can be rewritten as
\begin{equation}\label{Add-5}
\begin{split}
     {\bf w}^H {\bf Q} {\bf w} + {\rm Tr}\{ {\bf q}{\bf q}^H \}\leq P_C,
\end{split}
\end{equation}
where ${\bf Q} = (P_A {\bf h}_A{\bf h}^H_A + P_B {\bf h}_B{\bf h}^H_B + \sigma^2_C {\bf I}_M)^T  \otimes {\bf I}_M$.
Based on \eqref{Add-1}, \eqref{Add-4} and \eqref{Add-5}, optimization problem \eqref{II-1-eq6} can be recast into the following form by introducing new variables ${\bf X}= {\bf q} {\bf q}^H$ and ${\bf Y}={\bf w} {\bf w}^H$
\begin{eqnarray} \label{Add-6}
      && \max_{{\bf X} \succeq 0,{\bf Y} \succeq 0 } ~ \frac{{\rm Tr}( {\bf Q}_{01} {\bf X} ) } {  {\rm Tr}( {\bf Q}_{02} {\bf Y})  +\sigma^2_D}   \\ \nonumber
      s.t. &&  {\rm Tr} \left( {\bf Q}^i_{13} {\bf Y} \right) -  \left( {\bf Q}^i_2 {\bf X} \right) \geq \sigma^2_i,~i=A,B \\ \nonumber
      && {\rm Tr} ({\bf Q} {\bf Y}) + {\rm Tr}\{ {\bf X} \} \leq P_C \\ \nonumber
     && {\rm Rank}({\bf X})=1,~{\rm Rank}({\bf Y})=1
\end{eqnarray}
where ${\bf Q}^i_{13} = \frac{1}{\tau_i} {\bf Q}^i_1 - {\bf Q}^i_3$.
Due to the rank-one constraints,
finding the optimal solution of \eqref{Add-6} is difficult.
We therefore resort to relaxing it
by deleting the rank-one constraints, namely,
\begin{subequations}\label{Add-7}
\begin{align}
      & \max_{{\bf X} \succeq 0,{\bf Y} \succeq 0 } ~ \frac{{\rm Tr}( {\bf Q}_{01} {\bf X} ) } {  {\rm Tr}( {\bf Q}_{02} {\bf Y})  +\sigma^2_D}  \label{Add-7-1} \\
      s.t. ~~&  {\rm Tr} \left( {\bf Q}^i_{13} {\bf Y} \right) -  \left( {\bf Q}^i_2 {\bf X} \right) \geq \sigma^2_i,~i=A,B \label{Add-7-2} \\
      & {\rm Tr} ({\bf Q} {\bf Y}) + {\rm Tr}\{ {\bf X} \} \leq P_C \label{Add-7-3}
\end{align}
\end{subequations}
Then we shall show that the optimal rank-one solution of \eqref{Add-6} can be obtained from the relaxed problem \eqref{Add-7}.
According to \cite{Boyd2004}, optimization problem \eqref{Add-7} is a quasi-convex problem due to the fractional structure of the objective function in \eqref{Add-7-1}. In general, optimization problem \eqref{Add-7} can be solved through bisection search, which however has high computational complexity. Here we develop an alternative way to solve \eqref{Add-7} by using the Charnes-Cooper transformation \cite{Charnes1962}. Let
\begin{equation}\label{Add-8} \nonumber
     z = \frac{ 1 } {  {\rm Tr}( {\bf Q}_{02} {\bf Y})  +\sigma^2_D}.
\end{equation}
By defining $\bar{{\bf X}}= z{\bf X}$ and $\bar{{\bf Y}}= z{\bf Y}$, we can rewrite \eqref{Add-7} as
\begin{eqnarray}\label{Add-9}
      && \max_{\bar{{\bf X}} \succeq 0,\bar{{\bf Y}} \succeq 0 , z} ~  {\rm Tr}( {\bf Q}_{01} \bar{{\bf X}} )     \\ \nonumber
      s.t. &&  {\rm Tr}( {\bf Q}_{02} \bar{{\bf Y}})  +z \sigma^2_D = 1  \\  \nonumber
      &&  {\rm Tr} \left( {\bf Q}^i_{13} \bar{{\bf Y}} \right) -  {\rm Tr} \left( {\bf Q}^i_2 \bar{{\bf X}} \right) \geq z\sigma^2_i,~i=A,B  \\ \nonumber
      && {\rm Tr} ({\bf Q} \bar{{\bf Y}}) + {\rm Tr}\{ \bar{{\bf X}} \} \leq z P_C
\end{eqnarray}
After the transformation, it is easy to verify that \eqref{Add-9} is a standard semidefinite programming problem, thus its optimal solution can be easily obtained \cite{CVX}. Suppose that the optimal solution of \eqref{Add-9} is $\{\bar{{\bf X}}^\star, \bar{{\bf Y}}^\star,  z^\star\}$, the optimal solution of \eqref{Add-7}, denoted by $\{ {\bf X}^\star, {\bf Y}^\star \}$, can always be obtained through ${\bf X}^\star=\frac{\bar{{\bf X}}^\star}{z^\star}$ and ${\bf Y}^\star=\frac{\bar{{\bf Y}}^\star}{z^\star}$.
It is worth noting that if
$\bar{{\bf X}}^\star$ and $\bar{{\bf Y}}^\star$ are rank-one, then the optimal solution of \eqref{Add-6} can be obtained by using eigenvalue decomposition. Otherwise, the optimal rank-one solution of \eqref{Add-9} can be derived from the following theorem.

\textbf{Theorem} 1: If $\bar{{\bf X}}^\star$ and $\bar{{\bf Y}}^\star$ have higher rank than one, the optimal rank-one solution of \eqref{Add-9} can be obtained by using the following procedure.

\vspace{0.1cm}
\par
{\footnotesize
\begin{itemize}
\item Let $r_X$ and $r_Y$ denote the ranks of $\bar{{\bf X}}^\star$ and $\bar{{\bf Y}}^\star$, respectively;
\item \textbf{Repeat}
\begin{itemize}
\item Decompose $\bar{{\bf X}}^\star$ as $\bar{{\bf X}}^\star={\bf V}_X{\bf V}^H_X$ with ${\bf V}_X \in {\mathbb{C}}^{ N\times r_X}$ and $\bar{{\bf Y}}^\star$ as $\bar{{\bf Y}}^\star={\bf V}_Y{\bf V}^H_Y$ with ${\bf V}_Y \in {\mathbb{C}}^{ M^2 \times r_Y}$;
\item Find a nonzero $r_X \times r_X$ Hermitian matrix ${\bf M}_X$ and a $r_Y \times r_Y$ Hermitian matrix ${\bf M}_Y$ to satisfy the following linear equations
\begin{equation}\label{App-B4} \nonumber
\begin{split}
       & {\rm Tr}( {\bf V}^H_Y {\bf Q}_{02} {\bf V}_Y {\bf M}_Y )   = 0 \\
       &  {\rm Tr} \left({\bf V}^H_Y {\bf Q}^i_{13} {\bf V}_Y {\bf M}_Y \right) -  \left( {\bf V}^H_X {\bf Q}^i_2 {\bf V}_X {\bf M}_X \right) =0,~i=A,B  \\
       & {\rm Tr} ({\bf V}^H_Y {\bf Q} {\bf V}_Y {\bf M}_Y) + {\rm Tr}\{{\bf V}^H_X  {\bf V}_X {\bf M}_X \} =0
\end{split}
\end{equation}
\item Evaluate the eigenvalues $\varrho_{X,1},\varrho_{X,2},\cdots,\varrho_{X,r_X}$ of ${\bf M}_X$ and set $|\varrho_X|=\max \{|\varrho_{X,i}|,\forall i\}$, and the eigenvalues $\varrho_{Y,1},\varrho_{Y,2},\cdots,\varrho_{Y,r_Y}$ of ${\bf M}_Y$ and set $|\varrho_Y|=\max \{|\varrho_{Y,i}|,\forall i\}$;
\item Generate new matrices as $\bar{{\bf X}}'={\bf V}_X \left({\bf I}_{r_X} -(1/\varrho_X){\bf M}_X \right){\bf V}^H_X $ and $\bar{{\bf Y}}'={\bf V}_Y \left({\bf I}_{r_Y} -(1/\varrho_Y){\bf M}_Y \right){\bf V}^H_Y $, and set $\bar{{\bf X}}^\star=\bar{{\bf X}}'$ and $\bar{{\bf Y}}^\star=\bar{{\bf Y}}'$;
\end{itemize}
\item \textbf{Until} the ranks of $\bar{{\bf X}}^\star$ and $\bar{{\bf Y}}^\star$ are both equal to $1$.
\end{itemize}}
\vspace{0.1cm}

\begin{proof}
The proof is based on \textit{Theorem 3.2} in \cite{Huang2010}. Since there are four constraints in optimization problem \eqref{Add-8}, by using the above procedure, to satisfy Eq. (24) in \cite{Huang2010}, the ranks of $\bar{{\bf X}}^\star$ and $\bar{{\bf Y}}^\star$ should be both equal to $1$.
The proof of \emph{Theorem 1} is thus completed.
\end{proof}
By acquiring the optimal rank-one solution of \eqref{Add-9}, we can further get the optimal solution of \eqref{Add-6} and then the optimal solution of \eqref{II-1-eq6}.

\subsection{Joint Design of ${\bf Q}_{AB}$ and ${\bf Q}_s$ Under DF-XOR Two-Way Relay Strategy}
In this subsection, we consider the case where the secondary node $C$ adopts the DF-XOR two-way relay strategy. We assume that $C$ has correctly decoded the received signals from the two PUs in the first phase. Otherwise, we claim that the primary transmission is in outage.
For this relay strategy, the successful and unsuccessful interference subtractions in \eqref{I-eq13} lead to different problem formulations. They are thus treated separately in what follows. Note again that for this relay strategy, only both the signals $s_A$ and $s_B$ are correctly decoded in the first phase at $D$, the interference term ${\bf g}^T_D  {\bf s}_{AB}$ can be completely subtracted from \eqref{I-eq13}.

Firstly, we assume that the secondary receiver cannot cancel the interference caused by the PUs in \eqref{I-eq13}.  The corresponding optimization problem is thus given by
\begin{subequations}\label{II-2-eq1}
\begin{align}
    & \max_{{\bf Q}_{AB}\succeq 0, {\bf Q}_s\succeq 0} ~\log_2 \left(1 + \frac{{\bf g}^T_D {\bf Q}_s {\bf g}^*_D}{ {\bf g}^T_D {\bf Q}_{AB} {\bf g}^*_D  + \sigma^2_D}\right) \label{II-2-eq1-1}\\
    s.t. ~& \frac{1}{2}\log_2 \left(1 + \frac{{\bf g}^T_i {\bf Q}_{AB} {\bf g}^*_i }{{\bf g}^T_i {\bf Q}_s {\bf g}^*_i+ \sigma^2_i }\right) \geq R,~i=A,B \label{II-2-eq1-2}\\
     & {\rm Tr}\left( {\bf Q}_{AB} \right) + {\rm Tr}\left( {\bf Q}_s \right) \leq P_C  \label{II-2-eq1-3}
\end{align}
\end{subequations}
where ${\bf Q}_{AB}$ and ${\bf Q}_s$ are the covariance matrices of ${\bf s}_{AB}$ and ${\bf s}$, respectively, as defined in \eqref{I-eq10-1}, and $R=\max\{R_A, R_B\}$. The constraint \eqref{II-2-eq1-2} indicates that the transmission rate of the XORed signal from the secondary node $C$ should be larger than the maximizer of $\{R_A, R_B\}$ such that both primary receivers can successfully decode the combined information. As in \emph{Lemma 1}, we can also prove that the optimal ${\bf Q}_{AB}$ and ${\bf Q}_s$ can be rank-one. By defining ${\bf Q}_{AB}=\bar{{\bf w}}\bar{{\bf w}}^H$ and ${\bf Q}_s=\bar{{\bf q}}\bar{{\bf q}}^H$, the simplified beamforming design problem is yielded as
\begin{eqnarray} \label{II-2-eq2}
    && \max_{\bar{{\bf w}}, \bar{{\bf q}}} ~ \frac{|{\bf g}^T_D \bar{{\bf q}}|^2}{ |{\bf g}^T_D \bar{{\bf w}}|^2  + \sigma^2_D} \\ \nonumber
    s.t. &&  \frac{|{\bf g}^T_i \bar{{\bf w}}|^2 }{|{\bf g}^T_i \bar{{\bf q}}|^2 + \sigma^2_i } \geq \gamma,~i=A,B \\ \nonumber
     && ||\bar{{\bf w}}||^2_2 + ||\bar{{\bf q}}||^2_2 \leq P_C
\end{eqnarray}
where $\gamma =2^{2 R}-1$. To proceed to solve \eqref{II-2-eq2}, we have the following lemma.

\textbf{Lemma} 2: The optimal solution of \eqref{II-2-eq2} can be obtained in the following two cases:

\begin{itemize}
  \item If the dimension of space $\mathbb{G}$ defined in \emph{Lemma 1}, $N$, is larger than $1$, the optimal beamformers in \eqref{II-2-eq2} have the form of $\bar{{\bf w}}={\bf U}{\bf w} $ and $\bar{{\bf q}}={\bf U}{\bf q} $, where the optimal ${\bf w} $ and ${\bf q} $ can be obtained by solving the following problem
      \begin{eqnarray} \label{II-2-eq3}
    && \max_{{\bf w}, {\bf q}} ~ \frac{|{\bf t}_D {\bf q}|^2}{ |{\bf t}_D {\bf w}|^2  + \sigma^2_D} \\ \nonumber
    s.t. &&  \frac{|{\bf t}_i {\bf w}|^2 }{|{\bf t}_i {\bf q}|^2 + \sigma^2_i } \geq \gamma,~i=A,B \\ \nonumber
     && ||{\bf w}||^2_2 + ||{\bf q}||^2_2 \leq P_C
\end{eqnarray}
where ${\bf t}_D$ and ${\bf t}_i$, for $i \in \{A,B\}$, are defined as in \eqref{II-1-eq6}. Then by transforming \eqref{II-2-eq3} into an SDP problem as in \eqref{Add-9}, problem \eqref{II-2-eq3} can be optimally solved as \eqref{II-1-eq6}.
  \item If $N=1$, i.e., ${\bf u}_1={\rm Span}\{{\bf g}^*_D, {\bf g}^*_A, {\bf g}^*_B \}$, the optimal $\bar{{\bf q}}$ and $\bar{{\bf w}}$ can be denoted in the form as
  \begin{equation} \label{II-2-eq4}
  \bar{{\bf q}}= \sqrt{q} {\bf u}_1, ~~\bar{{\bf w}}=\sqrt{w} {\bf u}_1,
  \end{equation}
where $q$ and $w$ are two real positive scalars given, respectively, by
\begin{equation} \label{II-2-eq5}
  q = \frac{P_C - \gamma d }{\gamma+1}, ~~w = \frac{\gamma(P_C - \gamma d) }{\gamma+1} +\gamma d,
  \end{equation}
  where $d = \max \{ \sigma^2_A/t_A, \sigma^2_B/t_B \}$ with $t_A = |{\bf g}^T_A {\bf u}_1|^2$ and $t_B = |{\bf g}^T_B {\bf u}_1|^2$.
\end{itemize}
\begin{proof}
Please refer to Appendix~\ref{prof_lemma2}.
\end{proof}

From \emph{Lemma 2}, we find that when the channels in BC phase are parallel, the beamforming design can be significantly simplified and the closed-form solution can be obtained.

Secondly, we consider the scenario where the interference has been subtracted from \eqref{I-eq13} under the condition that both $s_A$ and $s_B$ have been correctly decoded in the first phase at $D$. The corresponding beamforming design problem can be written as
\begin{eqnarray} \label{II-2-eq6}
    && \max_{\bar{{\bf w}}, \bar{{\bf q}}} ~|{\bf g}^T_D \bar{{\bf q}}|^2 \\ \nonumber
    s.t. && \frac{|{\bf g}^T_i \bar{{\bf w}}|^2 }{|{\bf g}^T_i \bar{{\bf q}}|^2 + \sigma^2_i } \geq \gamma,~i=A,B \\ \nonumber
     && ||\bar{{\bf w}}||^2_2 + ||\bar{{\bf q}}||^2_2 \leq P_C
\end{eqnarray}
Although \eqref{II-2-eq6} has a simpler form than \eqref{II-2-eq2}, we can easily verify that \eqref{II-2-eq6} is still non-convex. In order to optimally solve \eqref{II-2-eq6}, we have the following lemma.

\textbf{Lemma} 3: The optimal solution of \eqref{II-2-eq6} can be obtained in the following three cases:

\begin{itemize}
  \item If we have orthonormal bases $\{{\bf u}_1, {\bf u}_2\}$ and $\{{\bf u}_1, \cdots, {\bf u}_N\}$ which satisfy ${\rm Span}\{{\bf u}_1, {\bf u}_2\} ={\rm Span}\{{\bf g}^*_A, {\bf g}^*_B\}$, and ${\rm Span}\{{\bf u}_1, \cdots, {\bf u}_N\}={\rm Span}\{{\bf g}^*_D, {\bf g}^*_A, {\bf g}^*_B\}$ with $N \geq 2$, the optimal $\bar{{\bf w}}$ and $\bar{{\bf q}}$ can be written in the form as
\begin{equation} \label{II-2-eq7}
\bar{{\bf w}} = w_A {\bf g}^*_A + w_B e^{-j\theta} {\bf g}^*_B, ~~\bar{{\bf q}} = \sum^N_{l=1} q_l {\bf u}_l,
\end{equation}
where $w_A$ and $w_B$ are two real positive scalars, $\theta = \angle {\bf g}^T_A {\bf g}^*_B$ and $q_l, \forall l$ are $N$ complex scalars.
By defining ${\bf w}=[w_A,w_B]^T$ and ${\bf q}=[q_1, \cdots, q_N]^T$, the optimal ${\bf w}$ and ${\bf q}$ can be obtained by solving the following second-order cone programming problem
\begin{eqnarray} \label{II-2-eq8}
    && \max_{{\bf w}, {\bf q}} ~ {\Re}({\bf t}_D {\bf q}) \\ \nonumber
    s.t.  && |{\bf t}_i {\bf q}|^2 + \sigma^2_i \leq \frac{1}{\gamma}({\bf a}_i {\bf w})^2,~i=A,B \\ \nonumber
    && || {\bf G}{\bf w}||^2_2 +  ||{\bf q}||^2_2 \leq P_C
\end{eqnarray}
where ${\bf G} = [{\bf g}^*_A, e^{-j\theta} {\bf g}^*_B]$,
${\bf t}_D$ and ${\bf t}_i$, for $i=\in \{A,B\}$, are defined as in \eqref{II-1-eq6},
${\bf a}_A = [||{\bf g}_A||^2_2, |{\bf g}^T_A {\bf g}^*_B|]$ and ${\bf a}_B = [|{\bf g}^T_B {\bf g}^*_A|, ||{\bf g}_B||^2_2]$.
  \item If ${\bf u}_1 ={\rm Span}\{{\bf g}^*_A, {\bf g}^*_B\}$ and ${\rm Span}\{{\bf u}_1, {\bf u}_2\}={\rm Span}\{{\bf g}^*_D, {\bf g}^*_A, {\bf g}^*_B\}$, the optimal $\bar{{\bf w}}$ and $\bar{{\bf q}}$ can be written in the form as
\begin{equation} \label{II-2-eq9}
\bar{{\bf w}} = \sqrt{w} {\bf u}_1, ~~\bar{{\bf q}} = \tilde{{\bf U}} {\bf q},
\end{equation}
where $w$ is a real positive scalar, $\tilde{{\bf U}}=[{\bf u}_1, {\bf u}_2]$ and ${\bf q}\in \mathbb{C}^2$. The optimal ${\bf q}$ and $w$ are given by
\begin{equation} \label{II-2-eq10}
{\bf q} = \sqrt{P_C-\gamma d} {\bf B}^{-\frac{1}{2}}{\bf v},~~w= \gamma d + \gamma |{\bf b} {\bf q}|^2,
\end{equation}
where $d=\max\{\sigma^2_A/|g_A|^2, \sigma^2_B/|g_B|^2 \} $ with $g_A$ and $g_B$ being defined as ${\bf g}^*_A = g_A {\bf u}_1$ and ${\bf g}^*_B = g_B {\bf u}_1$, ${\bf b} = {\bf u}^T_1 \tilde{{\bf U}}$, ${\bf v}$ is the eigenvector of ${\bf B}^{-\frac{1}{2}} {\bf A} {\bf B}^{-\frac{1}{2}}$ related to the maximum eigenvalue, and with ${\bf A}=\tilde{{\bf U}}^H {\bf g}^*_D {\bf g}^T_D \tilde{{\bf U}}$ and ${\bf B}=\gamma  {\bf b}^H {\bf b} +{\bf I}_2$.
   \item If ${\bf u}_1={\rm Span}\{{\bf g}^*_D, {\bf g}^*_A, {\bf g}^*_B\}$, the optimal $\bar{{\bf w}}$ and $\bar{{\bf q}}$ are given as in \eqref{II-2-eq4}.
\end{itemize}
\begin{proof}
Please refer to Appendix~\ref{prof_lemma3}.
\end{proof}

From \emph{Lemma 3}, we find that in general, when the interference is canceled at the secondary receiver $D$, the beamforming design can be simplified by recasting it into an SOCP problem, which can be solved more efficiently than the previous SDP problem. Similar to \emph{Lemma 2}, when the channels $\{{\bf g}^*_A, {\bf g}^*_B \}$ or $\{{\bf g}^*_D, {\bf g}^*_A, {\bf g}^*_B \}$ are parallel, the closed-form solution can be obtained.

\subsection{Joint Design of ${\bf Q}_A$, ${\bf Q}_B$ and ${\bf Q}_s$ Under DF-SUP Two-Way Relay Strategy}
In this subsection, we consider that the DF-SUP relay strategy is adopted at the secondary node $C$. In what follows, we also assume that $C$ has  perfectly recovered the information transmitted from the two PUs. Otherwise we claim that the primary transmission is in outage. Similar to the DF-XOR case, different formulations have been presented for with and without interference cancelation at the secondary receiver $D$.

Firstly, we consider the scenario where none of the interference terms has been subtracted from the received signal $y_{D,2}$ in \eqref{I-eq16}. Based on \eqref{I-eq13-1}, \eqref{I-eq14} and \eqref{I-eq16}, the optimization problem is formulated as
\begin{subequations}\label{II-3-eq1}
\begin{align}
    & \max_{ {\bf Q}_A, {\bf Q}_B, {\bf Q}_s } ~\log_2 \left(1 + \frac{{\bf g}^T_D {\bf Q}_s {\bf g}^*_D }{ {\bf g}^T_D {\bf Q}_A {\bf g}^*_D   +{\bf g}^T_D {\bf Q}_B {\bf g}^*_D + \sigma^2_D}\right) \label{II-3-eq1-1}\\
    s.t. ~& \frac{1}{2}\log_2 \left(1 + \frac{{\bf g}^T_i {\bf Q}_{\bar{i}} {\bf g}^*_i}{{\bf g}^T_i {\bf Q}_s {\bf g}^*_i+ \sigma^2_i }\right) \geq R_{\bar{i}},~i=A,B \label{II-3-eq1-2}\\
     & {\rm Tr}({\bf Q}_A) + {\rm Tr}({\bf Q}_B)+{\rm Tr}({\bf Q}_s) \leq P_C \label{II-3-eq1-3}
\end{align}
\end{subequations}
where ${\bf Q}_A$ and ${\bf Q}_B$ are the covariance matrices of ${\bf s}_A$ and ${\bf s}_B$, respectively, as defined in \eqref{I-eq13-1}.
Note that in constraint \eqref{II-3-eq1-2}, the rate thresholds for $s_A$ and $s_B$ are different since the messages to the two primary receivers are encoded separately.
Similar to the DF-XOR relay strategy, the optimal ${\bf Q}_i$, for $i\in \{A,B,s\}$, in \eqref{II-3-eq1} can be rank-one. Thus, by letting ${\bf Q}_A = \bar{{\bf w}}_A \bar{{\bf w}}^H_A$,
${\bf Q}_B = \bar{{\bf w}}_B \bar{{\bf w}}^H_B$ and ${\bf Q}_s = \bar{{\bf q}} \bar{{\bf q}}^H$, problem \eqref{II-3-eq1} is simplified as
\begin{eqnarray} \label{II-3-eq2}
    && \max_{ \bar{{\bf w}}_A, \bar{{\bf w}}_B, \bar{{\bf q}} } ~ \frac{|{\bf g}^T_D \bar{{\bf q}}|^2 }{ |{\bf g}^T_D \bar{{\bf w}}_A|^2   +|{\bf g}^T_D \bar{{\bf w}}_B|^2 + \sigma^2_D} \\ \nonumber
    s.t.~   &&  \frac{|{\bf g}^T_i \bar{{\bf w}}_{\bar{i}}|^2 }{|{\bf g}^T_i \bar{{\bf q}}|^2 + \sigma^2_i } \geq \tau_i,~i=A,B \\ \nonumber
     && ||\bar{{\bf w}}_A||^2_2 + ||\bar{{\bf w}}_B||^2_2 + ||\bar{{\bf q}}||^2_2 \leq P_C
\end{eqnarray}
where $\tau_i = 2^{2 R_{\bar{i}}}-1$, for $i\in \{A,B\}$, as defined in \eqref{II-1-eq5}. The optimal solution of \eqref{II-3-eq2} is summarized in the following lemma.

\textbf{Lemma} 4: With ${\rm Span}\{{\bf u}_1, \cdots, {\bf u}_N \}= {\rm Span}\{ {\bf g}^*_D, {\bf g}^*_A, {\bf g}^*_B\}$ and ${\bf U}=[{\bf u}_1, \cdots, {\bf u}_N]$ as in \emph{Lemma 1}, the optimal solution of \eqref{II-3-eq2} can be obtained in the following two cases:

\begin{itemize}
  \item When $N \geq 2$, the optimal solution of \eqref{II-3-eq2} can be written in the form as
  \begin{equation}\label{II-3-eq2-1}
  \bar{{\bf w}}_A = {\bf U}{\bf w}_A,~\bar{{\bf w}}_B = {\bf U}{\bf w}_B,~\bar{{\bf q}} = {\bf U}{\bf q},
  \end{equation}
  where the optimal ${\bf w}_A$, ${\bf w}_B$ and ${\bf q}$ can be obtained as in \eqref{II-1-eq6} by transforming \eqref{II-3-eq2} into an SDP problem as \eqref{Add-9}.
  \item When $N =1$, i.e., ${\bf u}_1 = {\rm Span}\{{\bf g}^*_D, {\bf g}^*_A, {\bf g}^*_B \}$, the optimal solution of \eqref{II-3-eq2} can be denoted in the form as
\begin{equation} \label{II-3-eq4}
\bar{{\bf q}} = \sqrt{q} {\bf u}_1,~~\bar{{\bf w}}_A = \sqrt{w_A} {\bf u}_1, ~~\bar{{\bf w}}_B = \sqrt{w_B} {\bf u}_1.
\end{equation}
The optimal coefficients in \eqref{II-3-eq4} are given, respectively, by
\begin{equation} \label{II-3-eq5}
q=\frac{P_C-\frac{\tau_B \sigma^2_B}{t_B}-\frac{\tau_A \sigma^2_A}{t_A}}{\tau_A + \tau_B + 1}, ~~w_A =q \tau_B +\frac{\tau_B \sigma^2_B}{t_B},~~w_B =q \tau_A +\frac{\tau_A \sigma^2_A}{t_A},
\end{equation}
where $t_i = |{\bf g}^T_i {\bf u}_1|^2$, for $i \in \{A,B\}$, as defined in \eqref{II-2-eq5}.
\end{itemize}
\begin{proof}
Please refer to Appendix~\ref{prof_lemma4}.
\end{proof}

Secondly, we consider the scenario where one of the interference terms has been subtracted from \eqref{I-eq16}. Without loss of generality, we assume that the signal ${\bf s}_A$ is canceled before demodulation, the corresponding beamforming design problem is formulated as
\begin{eqnarray} \label{II-3-eq6}
    && \max_{ \bar{{\bf w}}_A, \bar{{\bf w}}_B, \bar{{\bf q}} } ~ \frac{|{\bf g}^T_D \bar{{\bf q}}|^2 }{ |{\bf g}^T_D \bar{{\bf w}}_B|^2 + \sigma^2_D} \\ \nonumber
    s.t.~   &&  \frac{|{\bf g}^T_i \bar{{\bf w}}_{\bar{i}}|^2 }{|{\bf g}^T_i \bar{{\bf q}}|^2 + \sigma^2_i } \geq \tau_i,~i=A,B \\ \nonumber
     && ||\bar{{\bf w}}_A||^2_2 + ||\bar{{\bf w}}_B||^2_2 + ||\bar{{\bf q}}||^2_2 \leq P_C
\end{eqnarray}
Since \eqref{II-3-eq6} has a similar form with \eqref{II-3-eq2}, the proposed method in \emph{Lemma 4} can also be applied to solve \eqref{II-3-eq6}. Namely, when $N\geq 2$, problem \eqref{II-3-eq6} should be solved by transforming it into an SDP problem. While if $N=1$, the closed-form solution of \eqref{II-3-eq6} is derived as in \eqref{II-3-eq4}, which implies that when $N=1$, the optimal beamformers are irrelevant to the left interference term.

Finally, we consider the scenario where the two interference terms are completely subtracted from \eqref{I-eq16}, which leads to the following optimization problem
\begin{eqnarray} \label{II-3-eq9}
    && \max_{ \bar{{\bf w}}_A, \bar{{\bf w}}_B, \bar{{\bf q}} } ~ |{\bf g}^T_D \bar{{\bf q}}|^2  \\ \nonumber
    s.t.~   &&  \frac{|{\bf g}^T_i \bar{{\bf w}}_{\bar{i}}|^2 }{|{\bf g}^T_i \bar{{\bf q}}|^2 + \sigma^2_i } \geq \tau_i,~i=A,B \\ \nonumber
     && ||\bar{{\bf w}}_A||^2_2 + ||\bar{{\bf w}}_B||^2_2 + ||\bar{{\bf q}}||^2_2 \leq P_C
\end{eqnarray}
The optimal solution of \eqref{II-3-eq9} is summarized in the following lemma.

\textbf{Lemma} 5:
The optimal solution of \eqref{II-3-eq9} can be obtained in the following two cases:

\begin{itemize}
  \item When $N \geq 2$, the optimal solution of \eqref{II-3-eq9} can be written in the form as
\begin{equation} \label{II-3-eq10}
    \bar{{\bf q}}={\bf U}{\bf q},~~\bar{{\bf w}}_A = \sqrt{w_A} {\bf g}^*_B, ~~\bar{{\bf w}}_B = \sqrt{w_B} {\bf g}^*_A,
\end{equation}
where ${\bf U}$ is defined as in \emph{Lemma 1}. The optimal coefficients in \eqref{II-3-eq10} are given, respectively, by
\begin{equation} \label{II-3-eq11}
{\bf q} = \sqrt{\tilde{P}_C } {\bf D}^{-\frac{1}{2}}{\bf v}, ~~w_A =  \frac{\tau_B }{||{\bf g}_B||^4_2} {\bf q}^H {\bf t}^H_B {\bf t}_B {\bf q} + \frac{\tau_B \sigma^2_B }{||{\bf g}_B||^4_2},~~w_B =  \frac{\tau_A }{||{\bf g}_A||^4_2} {\bf q}^H {\bf t}^H_A {\bf t}_A {\bf q} + \frac{\tau_A \sigma^2_A }{||{\bf g}_A||^4_2},
\end{equation}
where $\tilde{P}_C = P_C - \frac{\tau_A \sigma^2_A}{||{\bf g}_A||^2_2}- \frac{\tau_B \sigma^2_B}{||{\bf g}_B||^2_2}$, ${\bf v}$ is the eigenvector of ${\bf D}^{-\frac{1}{2}} {\bf C} {\bf D}^{-\frac{1}{2}}$ related to the maximum eigenvalue, and with ${\bf C}={\bf t}^H_D {\bf t}_D$ and ${\bf D}= \frac{\tau_A }{||{\bf g}_A||^2_2}{\bf t}^H_A {\bf t}_A + \frac{\tau_B }{||{\bf g}_B||^2_2}{\bf t}^H_B {\bf t}_B + {\bf I}_N$.
  \item When $N=1$, i.e., ${\bf u}_1 ={\rm Span}\{{\bf g}^*_D, {\bf g}^*_A, {\bf g}^*_B \}$, the optimal solution of \eqref{II-3-eq9} can be written in the form as
\begin{equation} \label{II-3-eq12}
    \bar{{\bf q}}=\sqrt{q}{\bf u}_1,~~\bar{{\bf w}}_A = \sqrt{w_A} {\bf g}^*_B, ~~\bar{{\bf w}}_B = \sqrt{w_B} {\bf g}^*_A.
\end{equation}
The optimal coefficients in \eqref{II-3-eq12} are given, respectively, by
\begin{equation} \label{II-3-eq11-1}
q=\frac{P_C-\frac{\tau_A \sigma^2_A}{||{\bf g}_A||^2_2}-\frac{\tau_B \sigma^2_B}{||{\bf g}_B||^2_2}}{\frac{\tau_A |{\bf g}^T_A {\bf u}_1|^2}{||{\bf g}_A||^2_2} +\frac{\tau_B |{\bf g}^T_B {\bf u}_1|^2}{||{\bf g}_B||^2_2}+1}, ~~w_A =\frac{q \tau_B |{\bf g}^T_B {\bf u}_1|^2}{||{\bf g}_B||^4_2} + \frac{\tau_B \sigma^2_B}{||{\bf g}_B||^4_2},~~w_B =\frac{q \tau_A |{\bf g}^T_A {\bf u}_1|^2}{||{\bf g}_A||^4_2} + \frac{\tau_A \sigma^2_A}{||{\bf g}_A||^4_2 }.
\end{equation}
\end{itemize}
\begin{proof}
Please refer to Appendix~\ref{prof_lemma5}.
\end{proof}

\section{Simulation results}
In this section, some examples are presented to evaluate the proposed transceiver designs. We assume that the fading in each link follows independent Rayleigh distribution and the channel gain on each link is modeled by the distance path loss model, given as $\alpha_{i,j} = c \cdot d^{-n}_{i,j}$, where $c$ is an attenuation constant set as $1$, $n$ is the path loss exponent and fixed at $3$, and $d_{i,j}$ denotes the distance between nodes $i$ and $j$. Without loss of generality, we suppose that $d_{A,B}=1$. The node $D$ is placed in the perpendicular bisector of link $A \rightarrow B$ and the vertical distance from $D$ to the link $A \rightarrow B$ is $0.5$. Thus we have
\begin{equation}\label{III-eq1}\nonumber
   d_{C,D}= \sqrt{(\frac{d_{A,B}}{2}-d_{A,C})^2 + {0.5}^2}.
\end{equation}
For simplicity, the noise powers at all the destination nodes are set to be the same, i.e., $\sigma^2_A = \sigma^2_B =\sigma^2_C = \sigma^2_D= \sigma^2=1$ and the transmit powers at the two PUs are set as $P_A = P_B =P=5$ dB. During the first phase, the secondary receiver $D$ applies the simple successive interference cancelation (SIC) decoding where the stronger signal is always decoded first, followed by the weaker signal.
We let the rate requirements of the two PUs be $R_A = \alpha R$ and $R_B = (1-\alpha) R$, where the target sum-rate $R$ is given by $R = K R_0$  with $R_0 = \frac{1}{2}\log_2(1+\frac{P d^2_{A,B}}{\sigma^2})$ and $K$ being any constant.
Throughout the simulation, the performance is evaluated in two-folds. The first one is the maximum achievable rate of the SU given that the rate requirements of both PUs are satisfied. The other one is the outage probability that the rate requirements of the two PUs cannot be fulfilled.

\begin{figure}[tbhp]
  \centering
  \subfigure[Average achievable rate of SU]{
    \label{fig1:a} 
    \includegraphics[scale=0.5]{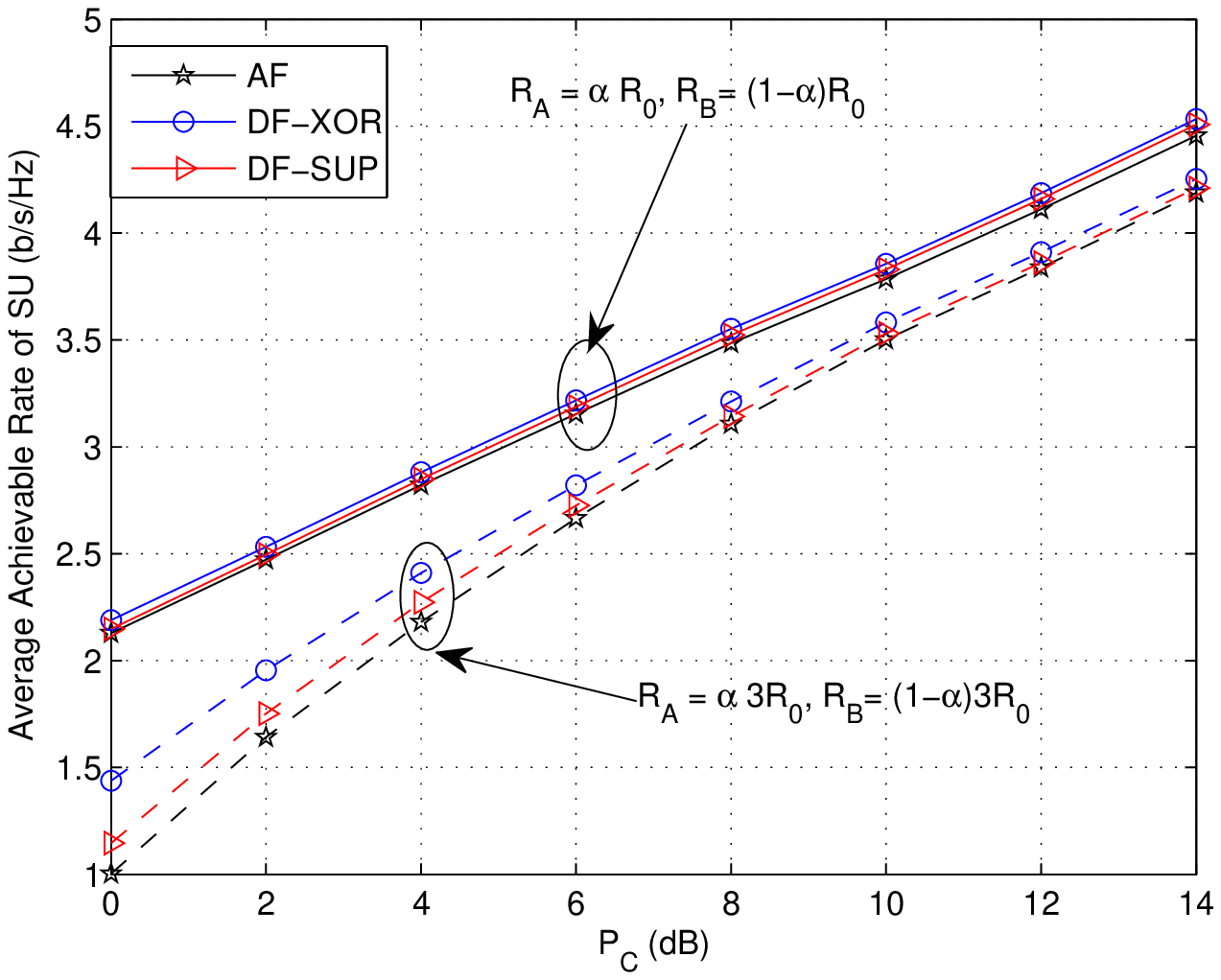}}
  \hspace{0in}
  \subfigure[Outage performance of primary transmission]{
    \label{fig1:b} 
    \includegraphics[scale=0.5]{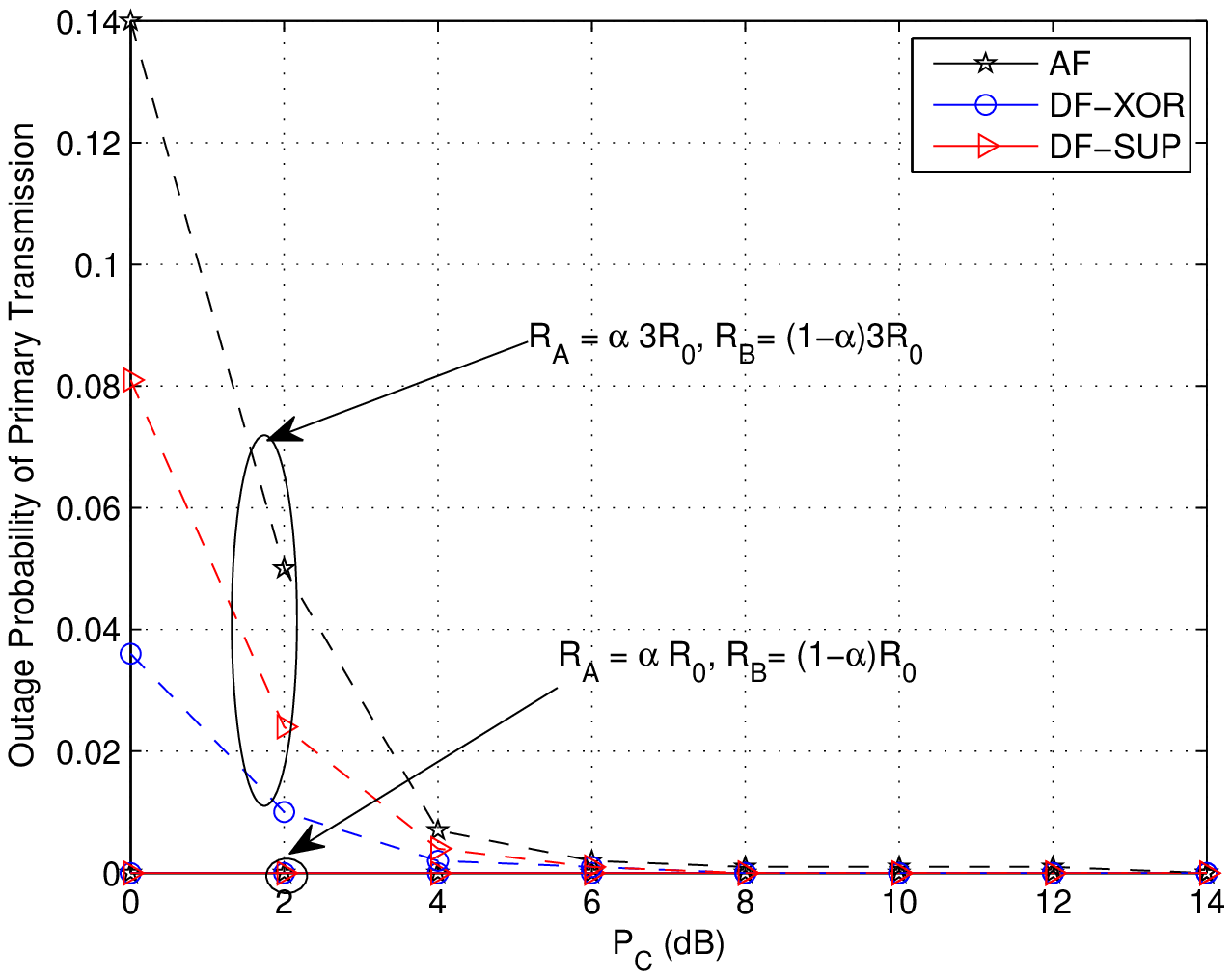}}
  \caption{Performance comparison for different relay strategies at $M=4$ with $\alpha=0.5$ when changing $P_C$.}
  \label{fig1} 
\end{figure}

\begin{figure}[tbhp]
  \centering
  \subfigure[Average achievable rate of SU]{
    \label{fig2:a} 
    \includegraphics[scale=0.5]{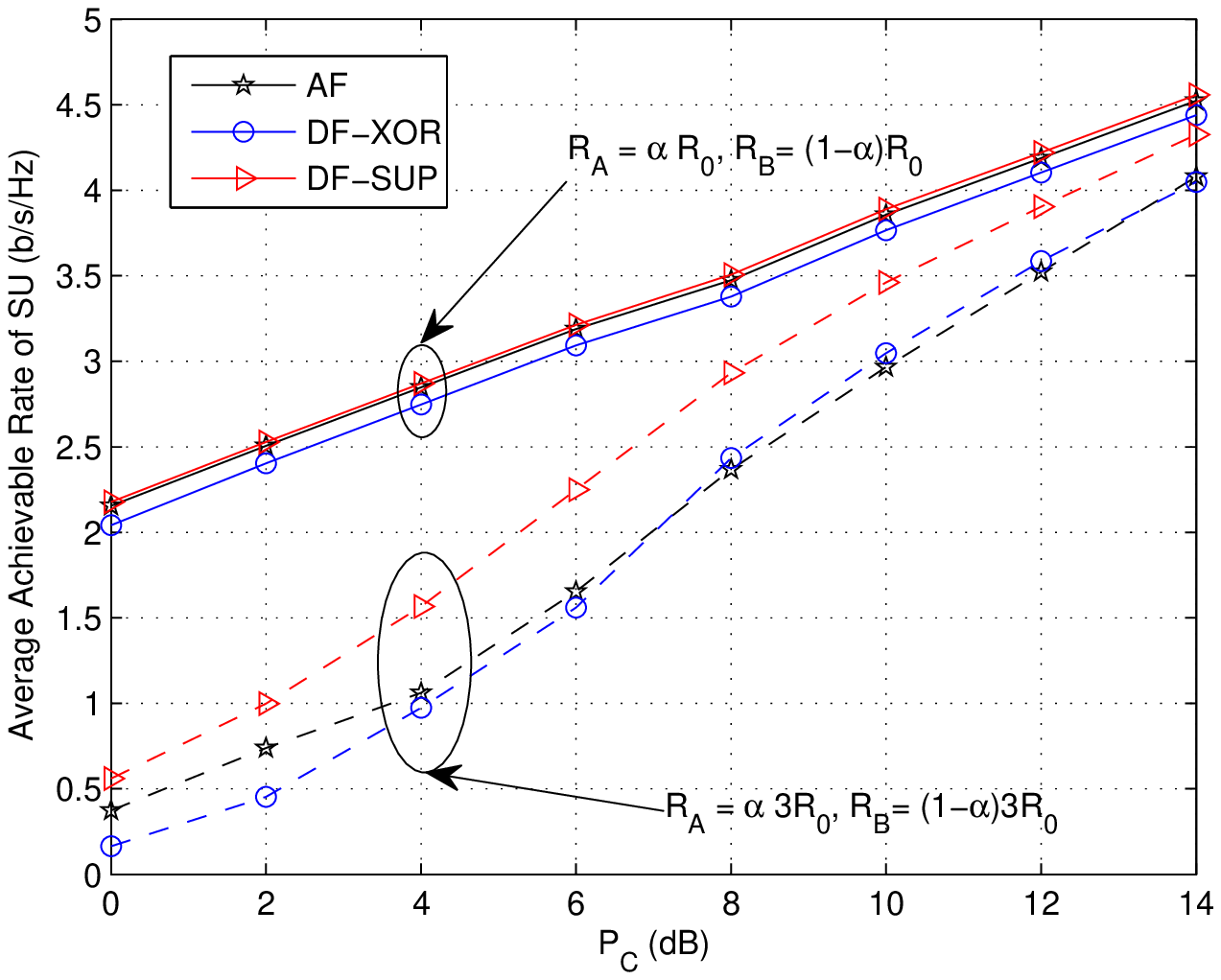}}
  \hspace{0in}
  \subfigure[Outage performance of primary transmission]{
    \label{fig2:b} 
    \includegraphics[scale=0.5]{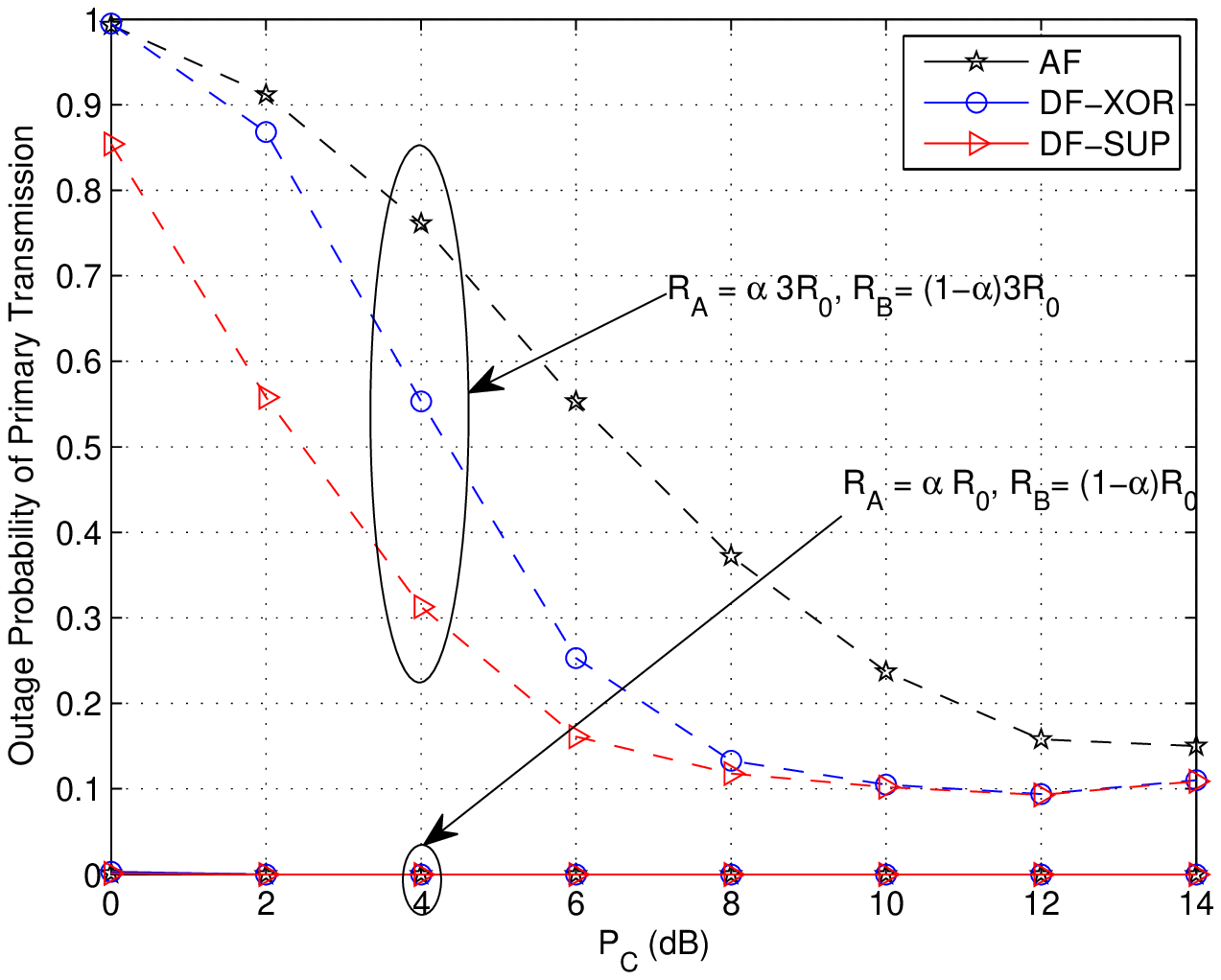}}
  \caption{Performance comparison for different relay strategies at $M=4$ with $\alpha=0.1$ when changing $P_C$.}
  \label{fig2} 
\end{figure}

In Fig.~\ref{fig1} and Fig.~\ref{fig2}, we illustrate the average achievable rate of the SU and the outage performance of the primary transmission in subfigures (a) and (b), respectively, as the function of the power $P_C$ by choosing $d_{A,C}=d_{B,C}=0.5$ and $M=4$. Specifically, the rate requirements of the two PUs are symmetric, i.e., $\alpha = 0.5$, in Fig.~\ref{fig1} and asymmetric with $\alpha = 0.1$ in Fig.~\ref{fig2}. For comparison, two different primary rate requirements with $K=1$ and $K=3$ are simulated for each scenario. From Fig.~\ref{fig1}, we find that
when the target sum-rate of the PUs is small ($R=R_0$), the three considered two-way relay strategies perform closely from both the primary and secondary user's perspectives. However, when the target primary sum-rate is high ($R=3R_0$),
the DF-XOR relay strategy performs the best, and the DF-SUP relay strategy outperforms the AF relay strategy. This indicates that under the symmetric scenario,
the secondary node $C$ would prefer to re-generate the primary signals when it wants to maximize the secondary transmission rate
since the destination noise at the secondary node $C$ is not accumulated for the subsequent transmission. Moreover, combining the information using XOR is better than using superposition
since the power of the secondary node $C$ can be used more efficiently in the DF-XOR relay strategy.
However, under the asymmetric condition, we observe from Fig.~\ref{fig2} that the DF-SUP relay strategy performs better than two other strategies, and the AF relay strategy begins to outperform the DF-XOR strategy when $K=1$. This is because when $R_A \neq R_B$, the bemaforming design for DF-XOR in \eqref{II-2-eq1} should make the achievable primary transmission rate larger than the maximizer of $R_A$ and $R_B$, which degrades the system performance.
While for the DF-SUP strategy, since different primary messages are encoded individually,
the power can be allocated to two primary messages more flexibly, which saves the power and improves the performance of the SU.
For the outage performance of the PU, we find that when the rate requirements of the PUs are small, i.e., $K=1$, the outage approaches zero for all the strategies. As the rate requirements increase, i.e., $K=3$, the outage of the primary transmission is increased significantly. In general, the AF relay strategy
has a higher outage probability
due to the accumulation of the back-propagated noise. In addition, the DF-SUP relay strategy has higher outage than the DF-XOR relay strategy under the symmetric primary rate requirements. While for the asymmetric case, the opposite result can be observed.

\begin{figure}[tbhp]
\begin{centering}
\includegraphics[scale=0.5]{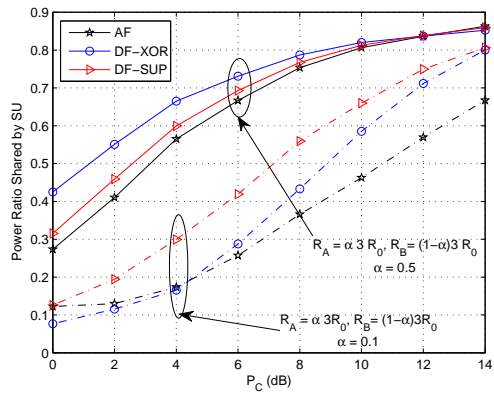}
\vspace{-0.1cm}
\caption{Power ratio shared by SU with target sum-rate $R=3 R_0$.} \label{SU_Ratio}
\end{centering}
\vspace{-0.3cm}
\end{figure}

In Fig.~\ref{SU_Ratio}, the power ratio shared by the SU, i.e., $\frac{{\rm Tr}({\bf Q}_s)}{P_C}$, is illustrated as the function of the secondary node power $P_C$ with target sum-rate $R=3 R_0$ at $d_{A,C}=d_{B,C}=0.5$ and $M=4$. We find that with symmetric primary rate requirements, the SU can share more power with the DF relay strategy than with the AF relay strategy. Moreover, the DF-XOR relay strategy needs less power to meet the primary rate requirements than the DF-SUP strategy. The observation is consistent with the comparison result given in Fig.~\ref{fig1:a}. While in the asymmetric scenario, as we explain earlier, the DF-XOR relay strategy needs more power than the DF-SUP relay strategy to satisfy the asymmetric primary rate requirements, which results in the performance degradation for the SU. It is noted that in asymmetric scenario, although the DF-XOR relay strategy can offer more power to the SU than the AF relay strategy as shown in Fig.~\ref{SU_Ratio}, the DF-XOR relay strategy still achieves close performance with the AF relay strategy as shown in Fig.~\ref{fig2:a}. The main reason is that with the DF-XOR relay strategy, the secondary receiver needs to correctly decode both primary signals simultaneously in the first phase to cancel the interference, which becomes difficult in the asymmetric primary transmission.

\begin{figure}[tbhp]
  \centering
  \subfigure[Symmetric case, $\alpha =0.5$]{
    \label{fig3:a} 
    \includegraphics[scale=0.5]{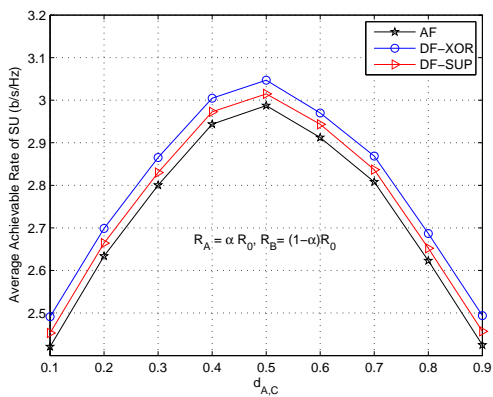}}
  \hspace{0in}
  \subfigure[Asymmetric case, $\alpha =0.1$]{
    \label{fig3:b} 
    \includegraphics[scale=0.5]{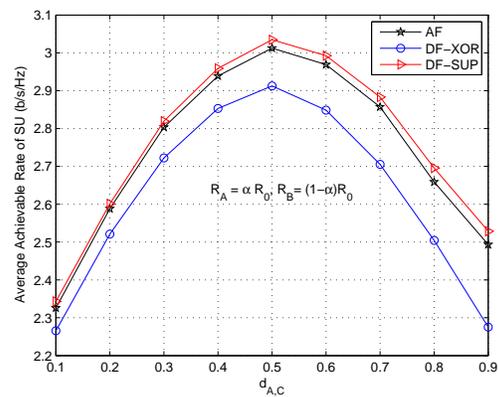}}
  \caption{Performance comparison for different relay strategies at $M=4$ and $P_C=5$ dB when changing $d_{A,C}$.}
  \label{fig3} 
\end{figure}

Fig.~\ref{fig3:a} and Fig.~\ref{fig3:b} illustrate the average achievable rate of the SU for symmetric primary rate requirements and asymmetric primary rate requirements, respectively, by changing distance $d_{A,C}$. The similar observations can be made as in Fig.~\ref{fig1} and Fig.~\ref{fig2}. Namely, under the symmetric primary rate requirements, the DF-XOR relay strategy performs the best, followed by the DF-SUP relay strategy and the AF relay strategy.
While under the asymmetric primary rate requirements, the DF-SUP relay strategy turns to perform the best,
and the AF relay strategy outperforms the DF-XOR relay strategy. From the plots, we find that all the strategies achieve the best performance at $d_{A,C}=0.5$ for both symmetric and asymmetric conditions. This implies that placing the secondary node $C$ in the middle of $A$ and $B$ is always the best choice.

\begin{figure}[tbhp]
  \centering
  \subfigure[Average achievable rate of SU]{
    \label{fig4:a} 
    \includegraphics[scale=0.5]{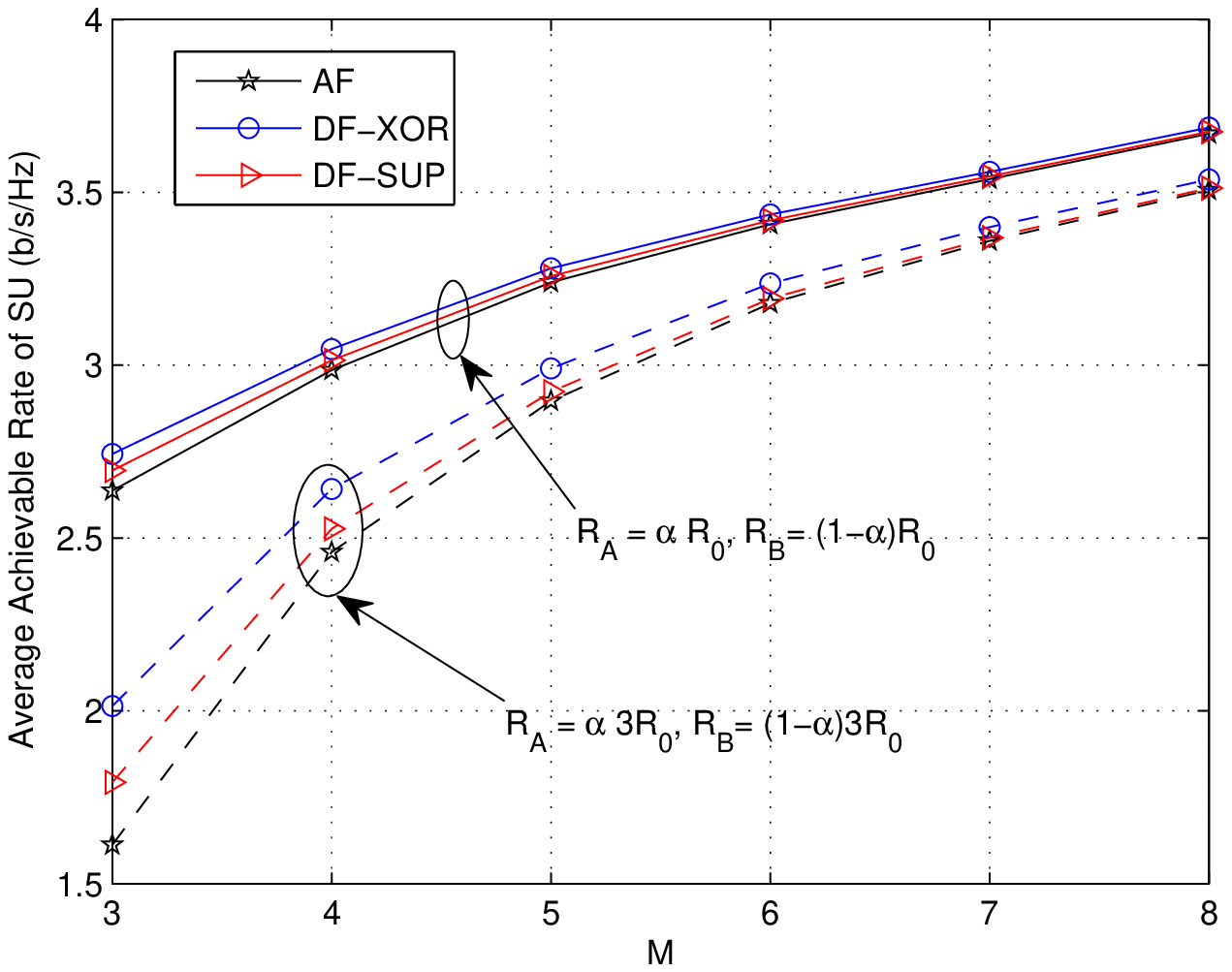}}
  \hspace{0in}
  \subfigure[Outage performance of primary transmission]{
    \label{fig4:b} 
    \includegraphics[scale=0.5]{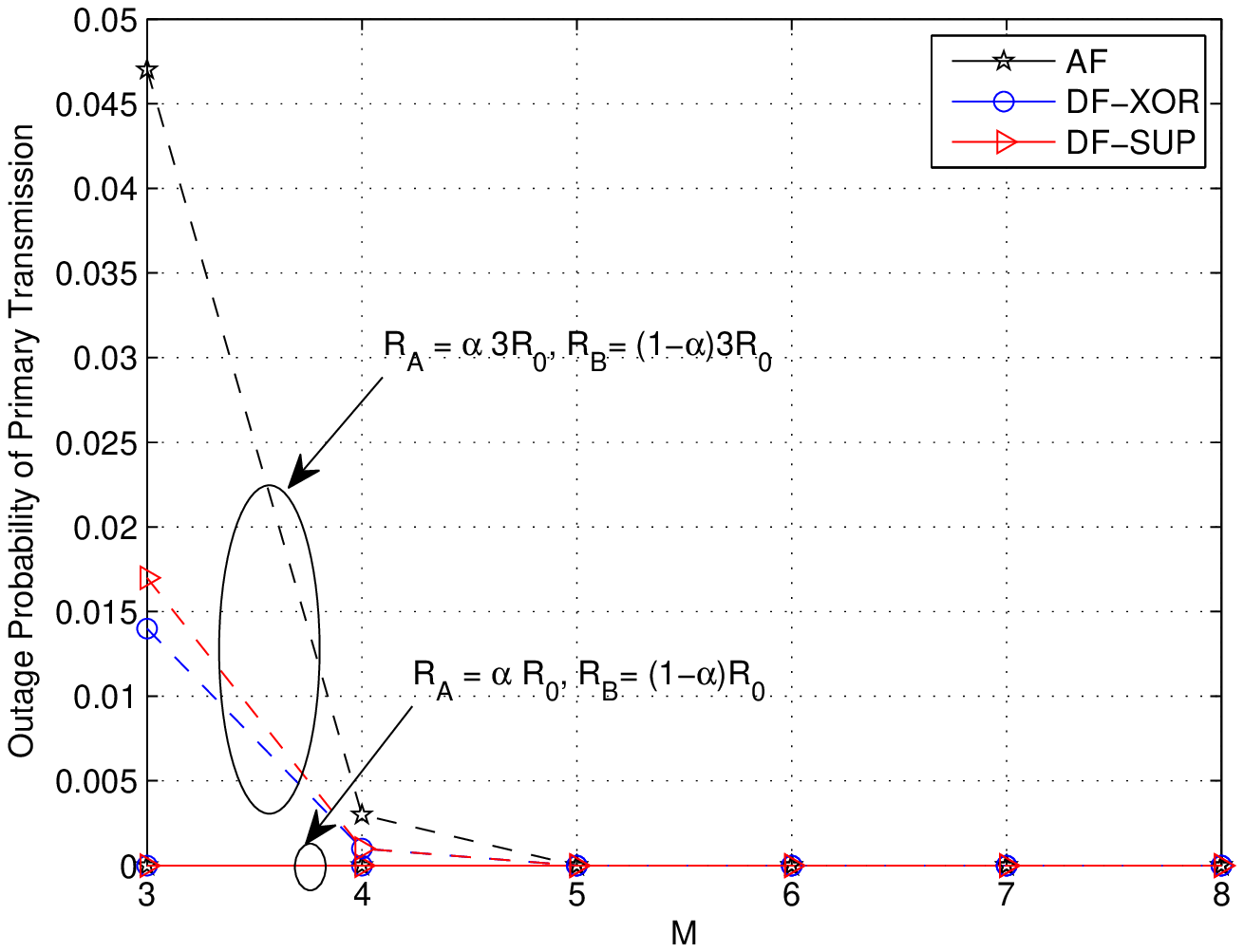}}
  \caption{Performance comparison for different relay strategies at $\alpha=0.5$ when changing $M$.}
  \label{fig4} 
\end{figure}

\begin{figure}[tbhp]
  \centering
  \subfigure[Average achievable rate of SU]{
    \label{fig5:a} 
    \includegraphics[scale=0.5]{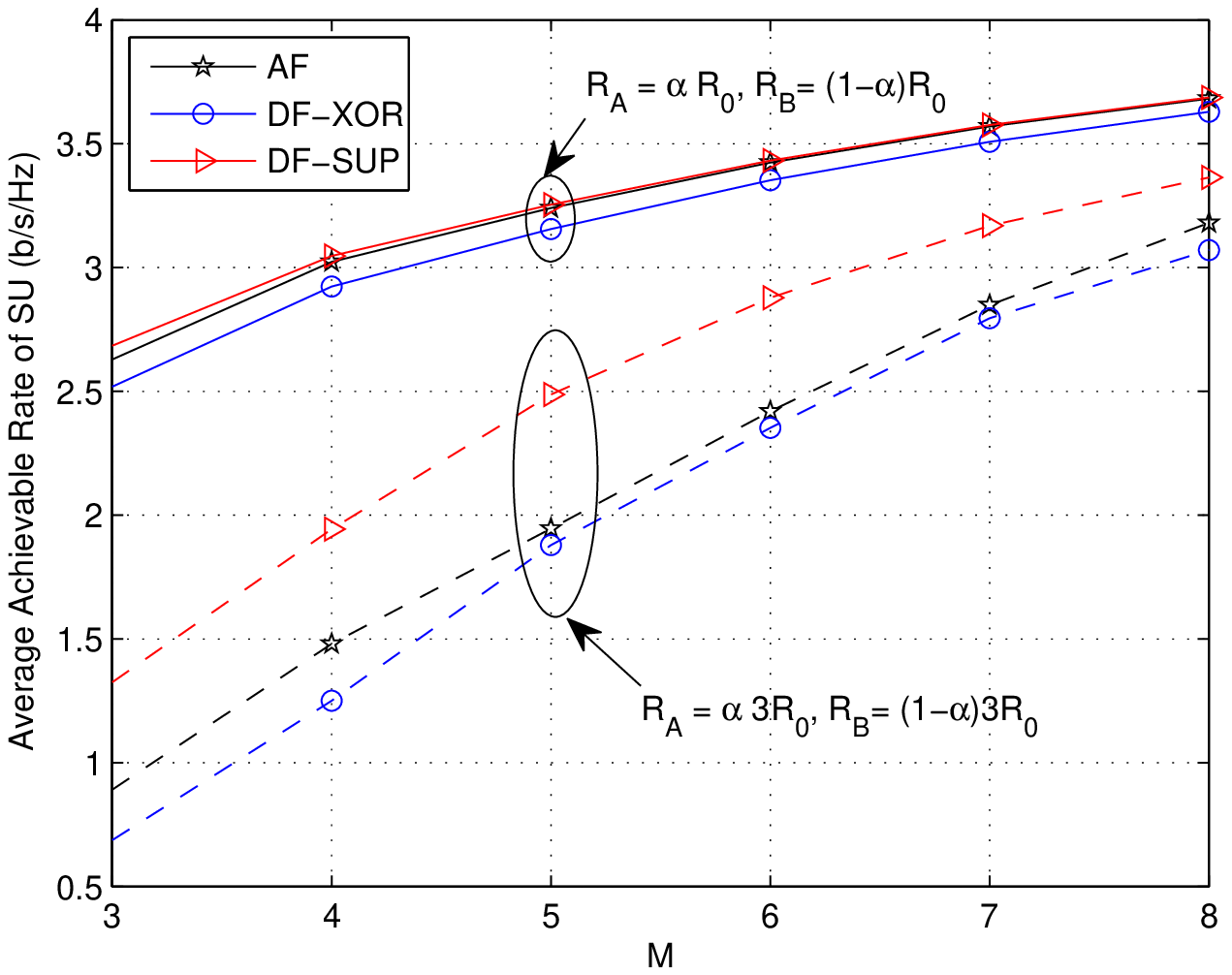}}
  \hspace{0in}
  \subfigure[Outage performance of primary transmission]{
    \label{fig5:b} 
    \includegraphics[scale=0.5]{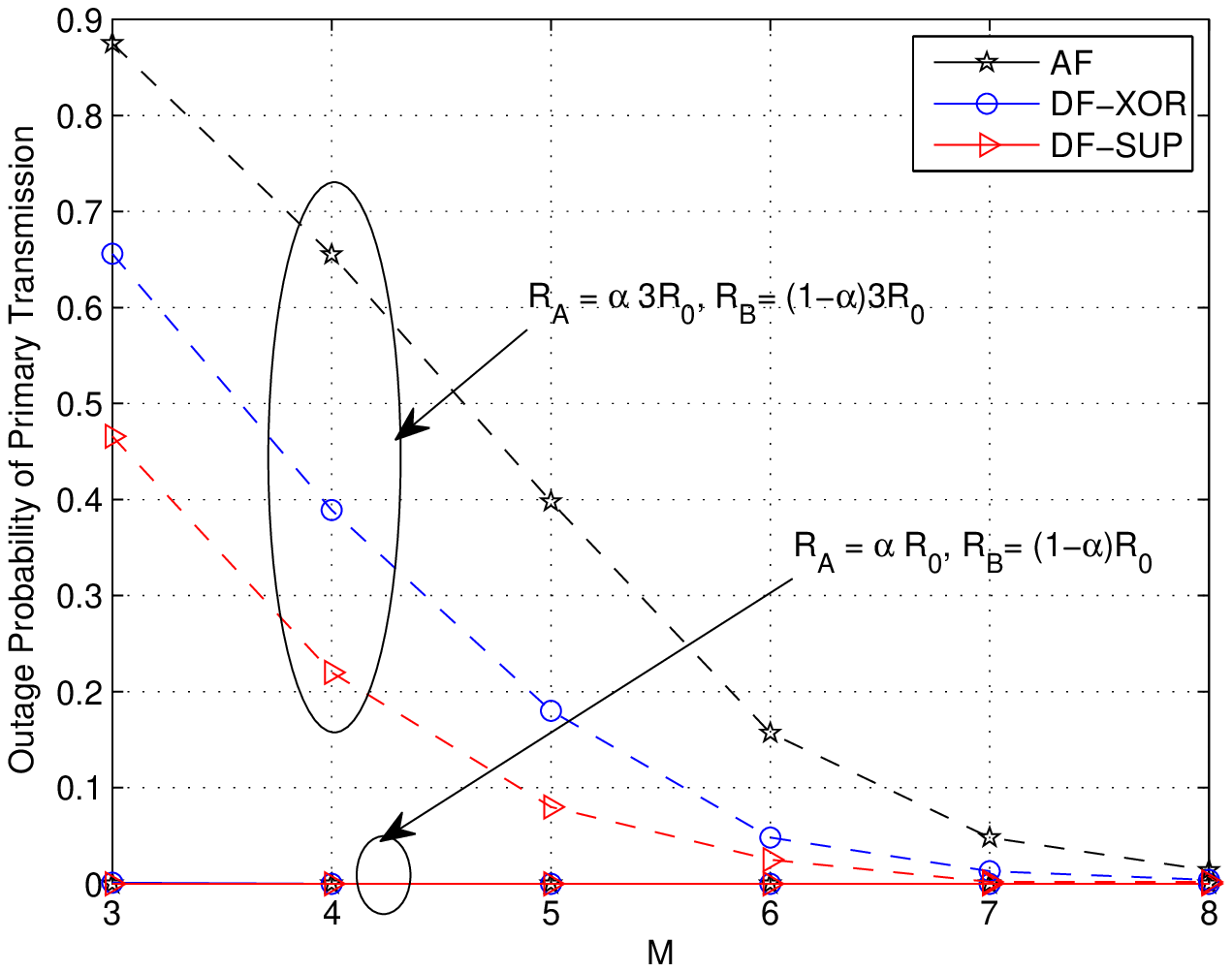}}
  \caption{Performance comparison for different relay strategies at $\alpha=0.1$ when changing $M$.}
  \label{fig5} 
\end{figure}

Finally, in Fig.~\ref{fig4} and Fig.~\ref{fig5}, the average achievable rate of the SU and the outage performance of the primary transmission are shown in subfigures (a) and (b), respectively, as the function of antenna number $M$ by setting $d_{A,C}=d_{B,C}=0.5$ and $P_C = 5$ dB. For the symmetric primary rate requirements in Fig.~\ref{fig4}, the similar comparison results can be observed as in Fig.~\ref{fig1}. Moreover, we find that as $M$ increases, the performance gap between three strategies becomes small and the outage for all the strategies approaches zero quickly. While for the asymmetric case in Fig.~\ref{fig5}, we find that when $K=1$, the DF-SUP relay strategy almost attains the same performance with the AF relay strategy when $M$ becomes large and they outperform the DF-XOR relay strategy. However, with larger primary rate requirements of $K=3$, we find that the DF-SUP relay strategy begins to significantly outperform the other two relay strategies, and the performance of the AF relay strategy is close to the DF-XOR relay strategy. While for the outage performance of the primary transmission with the asymmetric rate requirements, the same result can be observed as in Fig.~\ref{fig2:b}.

\section{Conclusions}
In this paper, we studied transceiver designs for a cognitive two-way relay network with the aim of maximizing the achievable transmission rate of the SU while maintaining the rate requirements of the PUs. Three different relay strategies were considered and the corresponding transceiver designs were formulated. By using suitable optimization tools, the optimal solutions were found for all the cases. Our simulation results showed that when the rate requirements of the two PUs are symmetric, the DF-XOR relay strategy performs the best and the least relay power is required to meet the rate requirements of the PUs. While the primary rate requirements are asymmetric, the DF-SUP performs the best along with the least relay power consumption to satisfy the rate requirement of the PUs.

\appendices
\section{Proof of lemma 1}
\label{prof_lemma1}
We assume that the optimal solution of ${\bf W}$ in \eqref{II-1-eq5} is denoted as $\bar{{\bf W}}$, then the optimal ${\bf Q}_s$ can be solved from the following optimization problem
\begin{eqnarray} \label{AppI-eq1}
      \max_{{\bf Q}_s\succeq 0 } && {\bf g}^T_D  {\bf Q}_s {\bf g}^*_D  \\ \nonumber
      s.t. && {\bf g}^T_i {\bf Q}_s {\bf g}^*_i \leq  o_i , ~i=A,B \\ \nonumber
     &&  {\rm Tr}({\bf Q}_s)  \leq \bar{P}_C
\end{eqnarray}
where $o_i = \frac{P_{\bar i} |{\bf g}^T_i \bar{{\bf W}} {\bf h}_{\bar i} |^2} {\tau_i}- \sigma^2_C ||{\bf g}^T_i \bar{{\bf W}} ||^2_2 -\sigma^2_i$ and $\bar{P}_C = P_C - (P_A ||\bar{{\bf W}} {\bf h}_A ||^2_2 + P_B||\bar{{\bf W}} {\bf h}_B ||^2_2 + {\sigma^2_C} ||\bar{{\bf W}}||^2_F)$. By using the circle property of trace operator, we can rewrite \eqref{AppI-eq1} as
\begin{eqnarray} \label{AppI-eq2}
      \max_{{\bf Q}_s\succeq 0 } &&   {\rm Tr} ({\bf C}_0 {\bf Q}_s)   \\ \nonumber
      s.t. && {\rm Tr} ( {\bf C}_i {\bf Q}_s ) \leq o_i, ~i=A,B \\ \nonumber
     &&  {\rm Tr}({\bf Q}_s)  \leq \bar{P}_C
\end{eqnarray}
where ${\bf C}_0 = {\bf g}^*_D {\bf g}^T_D$ and ${\bf C}_i = {\bf g}^*_i {\bf g}^T_i$. Then for \eqref{AppI-eq2}, we can use the same method as in the proof of \emph{Theorem 3.2} in \cite{Huang2010} to prove that the optimal ${\bf Q}_s$ can be rank-one although \eqref{AppI-eq2} has a different objective function from \cite{Huang2010}. To proceed, we first write the Lagrangian function of \eqref{AppI-eq2} as $\mathcal{L} = {\rm Tr} ({\bf C}_0 {\bf Q}_s) - \lambda_1 ({\rm Tr} ( {\bf C}_A {\bf Q}_s ) - o_A) - \lambda_2 ({\rm Tr} ( {\bf C}_B {\bf Q}_s ) - o_B) - \lambda_3 ({\rm Tr} (  {\bf Q}_s ) - \bar{P}_C)$ where $\lambda_i \geq 0$, for $i\in \{1,2,3\}$, are three Lagrangian multipliers. The corresponding Lagrangian dual function is yielded as $g(\lambda_1,\lambda_2,\lambda_3) = \sup_{{\bf Q}_s \succeq 0} ~ {\rm Tr}\{({\bf C}_0- \lambda_1 {\bf C}_A -\lambda_2 {\bf C}_B -\lambda_3 {\bf I}_M){\bf Q}_s\} + \lambda_1 o_A + \lambda_2 o_B + \lambda_3 \bar{P}_C$. The dual problem of \eqref{AppI-eq2} is thus written as
\begin{eqnarray} \label{AppI-eq3} \nonumber
      \min_{\lambda_1 \geq 0, \lambda_2 \geq 0,\lambda_3 \geq 0 } &&   \lambda_1 o_A + \lambda_2 o_B + \lambda_3 \bar{P}_C   \\ \nonumber
      s.t. && {\bf C}_0- \lambda_1 {\bf C}_A -\lambda_2 {\bf C}_B -\lambda_3 {\bf I}_M \preceq 0
\end{eqnarray}
Other than satisfying the constraints in \eqref{AppI-eq2}, the optimal solution of \eqref{AppI-eq2} should also satisfy the following complementary slackness conditions
\begin{equation}\label{AppI-eq4}
\begin{split}
    &\lambda_1 \left({\rm Tr} ( {\bf C}_A {\bf Q}_s ) - o_A\right) =0, \\
    &\lambda_2 \left({\rm Tr} ( {\bf C}_B {\bf Q}_s ) - o_B\right) =0, \\
    & \lambda_3 \left({\rm Tr} ( {\bf Q}_s ) - \bar{P}_C\right) =0, \\
    & {\rm Tr} \left( ({\bf C}_0- \lambda_1 {\bf C}_A -\lambda_2 {\bf C}_B -\lambda_3 {\bf I}_M){\bf Q}_s \right)=0.
\end{split}
\end{equation}
Since the number of the constraints in \eqref{AppI-eq2} is three, we can always apply the similar procedure provided in \emph{Algorithm 1} in \cite{Huang2010} to obtain a feasible rank-one solution to satisfy the conditions given in \eqref{AppI-eq4}. The brief proof is given as follows: suppose that the rank of the obtained ${\bf Q}_s$ in \eqref{AppI-eq2} is $r$ and it can be decomposed as ${\bf Q}_s={\bf V}{\bf V}^H$ with ${\bf V}\in {\mathbb C}^{M \times r}$.
Then a Hermitian matrix ${\bf M}$ is introduced to satisfy
\begin{equation}\label{AppI-eq4-1}
       {\bf Tr}\left({\bf V}^H {\bf C}_i {\bf V} {\bf M}\right)=0,~{\bf Tr}\left({\bf V}^H {\bf V} {\bf M}\right)=0,~i=A,B.
\end{equation}
If $r^2\geq 3$, we can always find a nonzero solution ${\bf M}$ satisfying \eqref{AppI-eq4-1}. By defining $\delta_i$, for $i\in \{1,2,\ldots,r\}$, as the eigenvalues of $\bf M$ and letting $|\delta_0|=\max \{|\delta_i|,\forall i\}$, we then get ${\bf Q}^{'}_s={\bf V}\left({\bf I}_R -(1/\delta_0){\bf M}\right){\bf V}^H$. It is easy to see that the rank of ${\bf Q}^{'}_s$ is reduced by at least one compared with ${\bf Q}_s$. In the meantime, we can check that ${\bf Q}^{'}_s$ is also a feasible solution of \eqref{AppI-eq2} and satisfies the optimal conditions in \eqref{AppI-eq4}, which further indicates that ${\bf Q}^{'}_s$ is also an optimal solution of \eqref{AppI-eq2} but with less rank than ${\bf Q}_s$. Repeat the above procedure until $r^2\leq 3$, an optimal rank-one solution of \eqref{AppI-eq2} is finally obtained.

Next we prove that the optimal beamformer regarding to ${\bf s}$ should lie in space $\mathbb{G}$ defined in \emph{Lemma 1}.
Note that the similar conclusion has been obtained for interference channel in \cite{Jorswieck2008}. We next give our proof with some differences.
By setting ${\bf Q}_s =\bar{{\bf q}} \bar{{\bf q}}^H$, problem \eqref{II-1-eq5} becomes
\begin{subequations}\label{AppI-eq5}
\begin{align}
    & \max_{{\bf W}, \bar{{\bf q}}} ~ \frac{|{\bf g}^T_D \bar{{\bf q}}|^2}{ a_A P_A|{\bf g}^T_D {\bf W} {\bf h}_A|^2 + a_B P_B|{\bf g}^T_D {\bf W} {\bf h}_B|^2 + \sigma^2_C ||{\bf g}^T_D {\bf W} ||^2_2 + \sigma^2_D} \label{AppI-eq5-1}\\
    s.t.~  &  \frac{P_{\bar i} |{\bf g}^T_i {\bf W} {\bf h}_{\bar i} |^2}{|{\bf g}^T_i \bar{{\bf q}}|^2 + \sigma^2_C ||{\bf g}^T_i {\bf W} ||^2_2 +\sigma^2_i } \geq \tau_i, ~i=A,B \label{AppI-eq5-2}\\
    & {\rm Tr}\left\{ {\bf W}(P_A {\bf h}_A{\bf h}^H_A + P_B {\bf h}_B{\bf h}^H_B + \sigma^2_C {\bf I}_M ){\bf W}^H \right\} + {\rm Tr}\left\{ \bar{{\bf q}}\bar{{\bf q}}^H\right\} \leq P_C \label{AppI-eq5-3}
\end{align}
\end{subequations}
It is assumed that space $\mathbb{C}^M$ is spanned by orthonormal bases $\{ {\bf u}_1,\cdots, {\bf u}_N, {\bf v}_1,\cdots,{\bf v}_{M-N} \}$ with ${\rm Span}\{{\bf u}_1,\cdots, {\bf u}_N \}={\rm Span}\{{\bf g}^*_D, {\bf g}^*_A, {\bf g}^*_B \}$. Without loss of generality, we assume that the optimal $\bar{{\bf q}}$ is given by $\bar{{\bf q}} = \sum^N_{l=1} \alpha_l {\bf u}_l + \sum^{M-N}_{l=1} \beta_l {\bf v}_l$ where $\alpha_l$ and $\beta_l$ are complex scalars.
It is easy to verify that the term $\beta_l {\bf v}_l$ does not affect the value of ${\bf g}^T_D \bar{{\bf q}}$, ${\bf g}^T_A \bar{{\bf q}}$ and ${\bf g}^T_B \bar{{\bf q}}$.
If there is a non-zero scalar $\beta_l$ which makes $\bar{{\bf q}}$ contain the vector ${\bf v}_l$, extra power of $P = \beta^2_l$ will be required. By denoting ${\bf T}=[{\bf g}^*_A, {\bf g}^*_B]$, we define $\Pi_g = {\bf T}({\bf T}^H {\bf T})^{-1} {\bf T}^H$ as the orthogonal projection onto space $\{{\bf g}^*_A, {\bf g}^*_B \}$ and $\Pi^{\bot}_g = {\bf I}-\Pi_g$ as the orthogonal projection onto the orthogonal complement of space $\{{\bf g}^*_A, {\bf g}^*_B \}$.
It is easy to verify that $\{\Pi_g {\bf g}^*_D ,  \Pi^{\bot}_g {\bf g}^*_D,  {\bf g}^*_A, {\bf g}^*_B \}$ spans the same space with $\{ {\bf g}^*_D,  {\bf g}^*_A, {\bf g}^*_B \}$. If we give the consumed extra power $P$ to the term $\Pi^{\bot}_g {\bf g}^*_D$ in $\bar{{\bf q}}$, we can always increase the value of the objective function while not affecting the constraints in \eqref{AppI-eq5}. This contradicts the optimality assumption made before.
Thus we complete the proof of \emph{Lemma 1}.

\section{Proof of lemma 2}
\label{prof_lemma2}
When $N \geq 2$, similar to \emph{Lemma 1}, it is easy to verify that optimization problem \eqref{II-2-eq2} can be simplified as \eqref{II-2-eq3}. Although \eqref{II-2-eq3} is a non-convex problem, by transforming it into an SDP problem, optimal solution can be obtained as in \eqref{Add-9}. Next we derive the optimal solution at $N=1$. Since ${\bf u}_1={\rm Span}\{{\bf g}^*_D, {\bf g}^*_A, {\bf g}^*_B \}$, according to \emph{Lemma 1}, the optimal $\bar{{\bf w}}$ and $\bar{{\bf q}}$ can be written in the form as
\begin{equation} \label{AppII-eq1}\nonumber
    \bar{{\bf q}}= \bar{q} {\bf u}_1, ~~\bar{{\bf w}}= \bar{w }{\bf u}_1,
\end{equation}
where $\bar{q}$ and $\bar{w }$ are two complex scalars. It is observed that multiplying $\bar{q}$ or $\bar{w }$ with an arbitrary phase shift does not affect the value of the objective function and the constraints in \eqref{II-2-eq2}. Thus, without loss of generality,
the optimal $\bar{{\bf w}}$ and $\bar{{\bf q}}$ can be written in the form as in \eqref{II-2-eq4}. By substituting \eqref{II-2-eq4} into \eqref{II-2-eq2}, problem \eqref{II-2-eq2} transforms into
\begin{eqnarray} \label{AppII-eq2}
    && \max_{ q, w} ~ \frac{q t_D}{ w t_D  + \sigma^2_D} \\ \nonumber
    s.t.~    && \frac{w t_i}{q t_i+ \sigma^2_i } \geq \gamma,~i=A,B\\ \nonumber
    && q+w \leq P_C
\end{eqnarray}
where $t_D= |{\bf g}^T_D {\bf u}_1|^2$ and $t_i$, for $i\in \{A,B\}$, is defined as in \eqref{II-2-eq5}. By defining
$d = \max \{ \sigma^2_A/t_A, \sigma^2_B/t_B \}$, problem \eqref{AppII-eq2} is equivalent to the following problem
\begin{subequations} \label{AppII-eq3}
\begin{align}
    & \max_{ q, w} ~ \frac{q t_D}{ w t_D  + \sigma^2_D} \label{AppII-eq3-1}\\
    s.t.  &~ \frac{w }{q + d } \geq \gamma \label{AppII-eq3-2} \\
    & q+w \leq P_C \label{AppII-eq3-3}
\end{align}
\end{subequations}
It is easy to observe that the optimal $q$ and $w$ in \eqref{AppII-eq3} must consume all the power to make constraint \eqref{AppII-eq3-3} active. Otherwise, the left power can always be assigned to $q$ and $w$ to further increase the value of the objective function, which contradicts the assumption of optimality. Besides that, we can also see that the optimal solution should make constraint \eqref{AppII-eq3-2} active. Otherwise, we can always lower $w$ to make the constraint \eqref{AppII-eq3-2} active and increase the value of the objective function. We thus acquire the following two equations
\begin{equation} \label{AppII-eq4} \nonumber
\frac{w }{q + d } = \gamma, ~~q+w = P_C.
\end{equation}
Then we obtain the optimal solution in \eqref{II-2-eq5}. The proof of \emph{Lemma 2} is thus completed.

\section{Proof of lemma 3}
\label{prof_lemma3}
It is observed that \eqref{II-2-eq6} have a similar form as\ \eqref{II-1-eq6}. Thus,
when $N \geq 2$, problem \eqref{II-2-eq6} can be solved by transforming it into an SDP problem similar to \eqref{Add-9} and then using \emph{Theorem 1} to obtain the optimal solution. Next we provide an alternative way to solve \eqref{II-2-eq6} by transforming it into an SOCP problem which can be solved more efficiently than the SDP problem. To conduct this transformation, we need to first prove the structure of the optimal beamformer given in \eqref{II-2-eq7}. For the optimal structure of $\bar{{\bf q}}$, the proof is similar to \emph{Lemma 1}. We next only focus on deriving the structure of optimal $\bar{{\bf w}}$.
Note that the similar form of beamformer has also been proven for the pure two-way relay channel in \cite{Oechtering2009}, we next show that it is also suitable to our considered case.
Since ${\rm Span}\{{\bf u}_1, {\bf u}_2 \}={\rm Span}\{{\bf g}^*_A, {\bf g}^*_B \}$, we can write the optimal $\bar{{\bf w}}$ in the form as $\bar{{\bf w}}=\bar{w}_A {\bf g}^*_A + \bar{w}_B {\bf g}^*_B$ with $\bar{w}_A$ and $\bar{w}_B$ being two complex scalars. Since any phase shift of $\bar{{\bf w}}$ does not affect its optimality, the optimal $\bar{{\bf w}}$ can be further denoted as
\begin{equation} \label{AppIII-eq1} \nonumber
\bar{{\bf w}} = w_A {\bf g}^*_A + w_B e^{j\phi}{\bf g}^*_B,
\end{equation}
where $w_A$ and $w_B$ are two real positive scalars. We assume that the optimal $\bar{{\bf w}}$ consumes the power of $P_W$ from $P_C$, i.e., $||\bar{{\bf w}}||^2_2=w^2_A ||{\bf g}_A||^2_2 + w^2_B ||{\bf g}_B||^2_2 + 2 w_A w_B |{\bf g}^T_A {\bf g}^*_B|\cos(\phi+\theta)=P_W$ with $\theta$ being defined in \eqref{II-2-eq7}. Then as in \cite{Oechtering2009}, the received signal power at the two primary receivers can be rewritten as
\begin{equation} \label{AppIII-eq2} \nonumber
\begin{split}
|{\bf g}^T_A \bar{{\bf w}}|^2 &=| w_A ||{\bf g}_A||^2_2 +w_B |{\bf g}^T_A {\bf g}^*_B|e^{j(\phi+\theta)}|^2 \\
& = ||{\bf g}_A||^2_2 \left(P_W -w^2_B (||{\bf g}_B||^2_2-\frac{|{\bf g}^T_A {\bf g}^*_B|^2}{||{\bf g}_A||^2_2}) \right)
\end{split}
\end{equation}
and $|{\bf g}^T_B \bar{{\bf w}}|^2 =||{\bf g}_B||^2_2 \left(P_W -w^2_A (||{\bf g}_A||^2_2-\frac{|{\bf g}^T_B {\bf g}^*_A|^2}{||{\bf g}_B||^2_2}) \right)$. We observe that if $\phi \neq -\theta$ at the optimal solution, we can always decrease the value of $w_A$ and $w_B$ to increase $|{\bf g}^T_A \bar{{\bf w}}|^2 $ and $|{\bf g}^T_B \bar{{\bf w}}|^2$ while keeping the consumed power $P_W$ constant. In this way, we can always extract some power from $\bar{{\bf w}}$ and give it to $\bar{{\bf q}}$ to increase the value of the objective function while keeping the constraints satisfied, which contradicts the assumption of optimality made before. We thus obtain \eqref{II-2-eq7}. Based on \eqref{II-2-eq7}, we have
\begin{equation} \label{AppIII-eq3}
\begin{split}
& |{\bf g}^T_A \bar{{\bf w}}|^2 = (w_A ||{\bf g}_A||^2_2 + w_B |{\bf g}^T_A {\bf g}^*_B|)^2=({\bf a}_A {\bf w})^2, \\
& |{\bf g}^T_B \bar{{\bf w}}|^2 = (w_A e^{-j\theta}|{\bf g}^T_B {\bf g}^*_A| + w_B e^{-j\theta} ||{\bf g}_B||^2_2) = ({\bf a}_B {\bf w})^2,
\end{split}
\end{equation}
where ${\bf a}_A$ and ${\bf a}_B$ are defined as in \eqref{II-2-eq8}. It is easy to see that both ${\bf a}_A {\bf w}$ and ${\bf a}_B {\bf w}$ in  \eqref{AppIII-eq3} are positive scalars. Moreover, using the structure of $\bar{{\bf q}}$ in \eqref{II-2-eq7}, we have
\begin{equation} \label{AppIII-eq4}
\begin{split}
 |{\bf g}^T_D \bar{{\bf q}}|^2 = |{\bf t}_D {\bf q}|^2, ~
 |{\bf g}^T_i \bar{{\bf q}}|^2 = |{\bf t}_i {\bf q}|^2,~~i=A,B.
\end{split}
\end{equation}
Note that in \eqref{AppIII-eq4}, for any optimal ${\bf q}$, we can always find a phase-shifted version $e^{j\vartheta}{\bf q}$ to make the scalar ${\bf t}_D {\bf q}$ real and positive while making $|{\bf t}_i \bar{{\bf q}}|^2$, for $i\in \{A,B\}$, constant. Thus, without loss of generality, we can maximize ${\Re}({\bf t}_D {\bf q})$ instead of $|{\bf t}_D {\bf q}|^2$ to get the optimal solution of
\eqref{II-2-eq6}, which leads to optimization problem \eqref{II-2-eq8}. It is easy to verify that \eqref{II-2-eq8} is a standard SOCP problem which can be efficiently solved \cite{CVX}.

When ${\bf u}_1 = {\rm Span}\{{\bf g}^*_A, {\bf g}^*_B \}$ and $\{ {\bf u}_1, {\bf u}_2 \}=\{ {\bf g}^*_D, {\bf g}^*_A, {\bf g}^*_B \}$, similar to \emph{Lemma 2}, we can prove that the optimal $\bar{{\bf w}}$ and $\bar{{\bf q}}$ in \eqref{II-2-eq6} have the form as in \eqref{II-2-eq9}. By assuming ${\bf g}^*_A = g_A {\bf u}_1$ and ${\bf g}^*_B = g_B {\bf u}_1$, problem \eqref{II-2-eq6} turns into
\begin{eqnarray} \label{AppIII-eq5}
    && \max_{w, {\bf q}} ~|{\bf a} {\bf q}|^2  \\ \nonumber
    s.t.~  && \frac{ w |g_i|^2}{|g_i|^2 |{\bf b} {\bf q}|^2+ \sigma^2_i } \geq \gamma,~i=A,B \\ \nonumber
     && ||{\bf q}||^2_2 + w \leq P_C
\end{eqnarray}
where ${\bf a}={\bf g}^T_D \tilde{{\bf U}}$ with $\tilde{{\bf U}}$ being defined in \eqref{II-2-eq9} and ${\bf b}$ is defined in \eqref{II-2-eq10}.
Problem \eqref{AppIII-eq5} is equivalent to the problem with the following form
\begin{subequations} \label{AppIII-eq6}
\begin{align}
    & \max_{w, {\bf q}} ~|{\bf a} {\bf q}|^2 \label{AppIII-eq6-1} \\
    s.t.~  & \frac{ w }{ |{\bf b} {\bf q}|^2+ d } \geq \gamma \label{AppIII-eq6-2}\\
     & ||{\bf q}||^2_2 + w \leq P_C \label{AppIII-eq6-3}
\end{align}
\end{subequations}
where $d$ is defined in \eqref{II-2-eq10}. It is seen that the optimal solution of \eqref{AppIII-eq6} must make constraint \eqref{AppIII-eq6-2} active, otherwise we can always extract some power from $w$ to make \eqref{AppIII-eq6-2} active and give it to ${\bf q}$ to further increase the value of the objective function. The active constraint \eqref{AppIII-eq6-2} leads to $w= \gamma d + \gamma |{\bf b} {\bf q}|^2$, which further simplifies \eqref{AppIII-eq6} as
\begin{eqnarray} \label{AppIII-eq7}
    && \max_{{\bf q}} ~|{\bf a} {\bf q}|^2  \\ \nonumber
    s.t.~   && ||{\bf q}||^2_2 + \gamma |{\bf b} {\bf q}|^2 \leq P_C - \gamma d
\end{eqnarray}
Problem \eqref{AppIII-eq7} can be rewritten as
\begin{eqnarray} \label{AppIII-eq8}
    && \max_{{\bf q}} ~{\bf q}^H {\bf A} {\bf q}  \\ \nonumber
    s.t.~   && {\bf q}^H {\bf B} {\bf q}  \leq P_C - \gamma d
\end{eqnarray}
where ${\bf A}$ and ${\bf B}$ are defined in \eqref{II-2-eq10}. By transforming \eqref{AppIII-eq8} into the following form
\begin{eqnarray} \label{AppIII-eq9}\nonumber
    && \max_{\tilde{{\bf q}}} ~\tilde{{\bf q}}^H {\bf B}^{-\frac{1}{2}} {\bf A} {\bf B}^{-\frac{1}{2}}\tilde{{\bf q}}  \\ \nonumber
    s.t.  && ||\tilde{{\bf q} } ||^2_2 = P_C - \gamma d
\end{eqnarray}
we thus obtain the solution given in \eqref{II-2-eq10}.

When ${\bf u}_1 ={\rm Span}\{{\bf g}^*_D, {\bf g}^*_A, {\bf g}^*_B\}$, the optimal solution can be obtained as in \emph{Lemma 2}. We then complete the proof of \emph{Lemma 3}.

\section{Proof of lemma 4}
\label{prof_lemma4}
As in \emph{Lemma 1}, the optimal $\bar{{\bf q}}$, $\bar{{\bf w}}_A$ and $\bar{{\bf w}}_B$ in \eqref{II-3-eq2} should have the form as in \eqref{II-3-eq2-1}. Then \eqref{II-3-eq2} can be simplified as
\begin{eqnarray} \label{AppIV-eq1}
    && \max_{ {\bf w}_A, {\bf w}_B, {\bf q} } ~ \frac{|{\bf t}_D {\bf q} |^2 }{ |{\bf t}_D {\bf w}_A|^2   +|{\bf t}_D {\bf w}_B|^2 + \sigma^2_D} \\ \nonumber
    s.t.~   &&  \frac{|{\bf t}_i {\bf w}_{\bar{i}}|^2 }{|{\bf t}_i {\bf q}|^2 + \sigma^2_i } \geq \tau_i,~i=A,B \\ \nonumber
     && ||{\bf w}_A||^2_2 + ||{\bf w}_B||^2_2 + ||{\bf q}||^2_2 \leq P_C
\end{eqnarray}
Similar to \eqref{II-1-eq6}, problem \eqref{AppIV-eq1} can be solved by transforming it into an SDP problem as \eqref{Add-9} and then the optimal solution is obtained by using \emph{Theorem 1}.

When $N=1$, similar to \emph{Lemma 2}, we obtain that the optimal solution should have the form given in \eqref{II-3-eq4}. Substituting them into \eqref{II-3-eq2}, we have
\begin{subequations} \label{AppIV-eq3}
\begin{align}
    & \max_{ q, w_A, w_B} ~ \frac{t_D q}{  t_D w_A + t_D w_B + \sigma^2_D} \label{AppIV-eq3-1}\\
    s.t.~  & q + w_A + w_B \leq P_C \label{AppIV-eq3-2}\\
    &  \frac{t_i w_{\bar{i}}}{t_i q+ \sigma^2_i } \geq \tau_i,~~i=A,B \label{AppIV-eq3-3}
\end{align}
\end{subequations}
where $t_D$ is defined in \eqref{AppII-eq2} and $t_i$ is defined in \eqref{II-3-eq5}.
In \eqref{AppIV-eq3}, we can verify that constraint \eqref{AppIV-eq3-3} must be active, otherwise we can always extract some power from $w_i$, for $i\in \{A,B\}$, to make constraint \eqref{AppIV-eq3-3} active and increase the value of the objective function. Hence we obtain the following two equations
\begin{equation}\label{AppIV-eq4} \nonumber
    \frac{t_A w_B}{t_A q+ \sigma^2_A } = \tau_A,~~\frac{t_B w_A}{t_B q+ \sigma^2_B } = \tau_B,
\end{equation}
which further lead to
\begin{equation}\label{AppIV-eq5}
   w_A = q \tau_B + \frac{\tau_B \sigma^2_B}{t_B},~~ w_B = q \tau_A + \frac{\tau_A \sigma^2_A}{t_A}.
\end{equation}
Since at the optimal solution, constraint \eqref{AppIV-eq3-2} should also be active. By substituting \eqref{AppIV-eq5} into \eqref{AppIV-eq3-2}, we obtain the optimal solution given in \eqref{II-3-eq5}.

\section{Proof of lemma 5}
\label{prof_lemma5}
Since in \eqref{II-3-eq9}, the beamformer $\bar{{\bf w}}_i$ is only related to the channel ${\bf g}_{{\bar i}}$, we therefore obtain that the optimal $\bar{{\bf w}}_i$ should have the form given in \eqref {II-3-eq10}. Substituting them into \eqref{II-3-eq9}, we have
\begin{subequations} \label{AppV-eq1}
\begin{align}
    & \max_{{\bf q}, w_A,w_B} ~{\bf q}^H {\bf t}^H_D {\bf t}_D {\bf q}  \label{AppV-eq1-1}\\
    s.t.~  &  \frac{w_{{\bar i}} ||{\bf g}_i||^4_2}{|{\bf t}_i {\bf q}|^2+ \sigma^2_i } \geq \tau_i,~i=A,B  \label{AppV-eq1-2}\\
     & w_A ||{\bf g}_B||^2_2 +  w_B ||{\bf g}_A||^2_2 + ||{\bf q}||^2_2 \leq P_C   \label{AppV-eq1-3}
\end{align}
\end{subequations}
Again using the fact that constraint \eqref{AppV-eq1-2} should be active, we obtain
\begin{equation} \label{AppV-eq2}
w_{{\bar i}} ||{\bf g}_i||^2_2 = \frac{\tau_i }{||{\bf g}_i||^2_2} {\bf q}^H {\bf t}^H_i {\bf t}_i {\bf q} + \frac{\tau_i \sigma^2_i }{||{\bf g}_i||^2_2},~i=A,B.
\end{equation}
Since the power constraint \eqref{AppV-eq1-3} should be active at the optimal solution, by combining with \eqref{AppV-eq2}, we have
\begin{eqnarray} \label{AppV-eq3} \nonumber
   && \max_{{\bf q}} ~ {\bf q}^H {\bf C} {\bf q} \\ \nonumber
    s.t.~   &&  {\bf q}^H {\bf D} {\bf q} = \tilde{P}_C
\end{eqnarray}
where $\tilde{P}_C$, ${\bf C}$, ${\bf D}$ are defined as in \eqref{II-3-eq11}. Similar to the proof of \emph{Lemma 3}, we finally obtain the optimal solution given in \eqref{II-3-eq11}.

When $N=1$, similar to \eqref{II-3-eq10}, the optimal beamformers should have the form given in \eqref{II-3-eq12}. Then we can simplify problem \eqref{II-3-eq9} by substituting them into \eqref{II-3-eq9}, which yields
\begin{eqnarray} \label{AppV-eq4}
    && \max_{q, w_A, w_B } ~q \\ \nonumber
    s.t.~  && \frac{w_{{\bar i}} ||{\bf g}_i ||^4_2}{q |{\bf g}^T_i {\bf u}_1|^2+ \sigma^2_i }\geq \tau_i,~i=A,B \\ \nonumber
    && w_A ||{\bf g}_B||^2_2 + w_B ||{\bf g}_A||^2_2 + q \leq P_C
\end{eqnarray}
Similar to the proof of \eqref{II-3-eq11}, we can derive the optimal coefficients given in \eqref{II-3-eq11-1} by using the fact that the optimal solution in \eqref{AppV-eq4} should make all the constraints active.

\bibliographystyle{IEEEtran}
\bibliography{IEEEabrv,Cognitive_TWR}

\end{document}